\documentclass{emulateapj}
\usepackage{lscape}

\shorttitle{A GLIMPSE of M16}
\shortauthors{R. Indebetouw et al.}

\newcommand{\rplotone}[1]{\plotone{#1}\\}

\newcommand{\um}{$\;\mu$m{} }
\newcommand{\msun}{M$_\sun${}}
\newcommand{\lsun}{L$_\sun${}}
\newcommand{\e}[1]{$\times$10$^{#1}${}}

\begin{document}

\title{Embedded Star Formation in the Eagle Nebula with {\it Spitzer}/GLIMPSE}
   
\author{R. Indebetouw\altaffilmark{1,2},
T. P. Robitaille\altaffilmark{3},
B. A. Whitney\altaffilmark{4},
E. Churchwell\altaffilmark{5}, B. Babler\altaffilmark{5}, M. Meade\altaffilmark{5},
C. Watson\altaffilmark{6}, M. Wolfire\altaffilmark{7}}

\altaffiltext{1}{University of Virginia, Astronomy Dept., P.O. Box 3818, Charlottesville, VA, 22903-0818}
\altaffiltext{2}{{\it Spitzer} fellow}
\altaffiltext{3}{SUPA, School of Physics and Astronomy, University of St. Andrews, North Haugh, St Andrews, KY16 9SS UK}
\altaffiltext{4}{Space Science Institute, 4750 Walnut St. Suite 205, Boulder, CO 80301}
\altaffiltext{5}{University of Wisconsin-Madison, Department of Astronomy, 475 North Charter Street, Madison, WI 53706}
\altaffiltext{6}{Manchester College, Dept. of Physics, North Manchester, IN, 46962}
\altaffiltext{7}{University of Maryland, Dept. of Astronomy, College Park, MD 20742-2421}

\begin{abstract}
We present new {\it Spitzer} photometry of the Eagle Nebula (M16, 
containing the optical cluster NGC 6611)
combined with near-infrared photometry from 2MASS.  We use dust
radiative transfer models, mid-infrared and near-infrared color-color
analysis, and mid-infrared spectral indices to analyze point source
spectral energy distributions, select candidate young stellar objects (YSOs), and
constrain their mass and evolutionary state.  Comparison of the
different protostellar selection methods shows that mid-infrared
methods are consistent, but as has been known for some time,
near-infrared-only analysis misses some young objects.  We reveal more
than 400 protostellar candidates, including one massive young stellar
object (YSO) that has not been previously highlighted.  The YSO
distribution supports a picture of distributed low-level star
formation, with no strong evidence of triggered star formation in the
``pillars''.  We confirm the youth of NGC~6611 by a large fraction of
infrared-excess sources, and reveal a younger cluster of YSOs in the
nearby molecular cloud.  Analysis of the YSO clustering properties
shows a possible imprint of the molecular cloud's Jeans length.
Multiwavelength mid-IR imaging thus allows us to analyze the
protostellar population, to measure the dust temperature and column
density, and to relate these in a consistent picture of star formation
in M16.
\end{abstract}


%%%%%%%%%%%%%%%%%%%%%%%%%%%%%%%%%%%%%%%%%%%%%%%%%%%%%%%%%%%%%%%%%%%

\section{Introduction}

The fusion energy generated in stars dominates the evolution of
galaxies, heating and processing their interstellar media.  However,
our understanding of how stars form is incomplete.  Some fairly
reliable empirical scaling laws exist, and a believable scenario
exists for how an isolated low-mass star can form via disk accretion,
but we lack a detailed understanding of how an entire molecular cloud
turns into a cluster of stars.  Most stars form in cluster
environments, and stellar densities imply some interaction between
YSOs during their formation.  We must therefore analyze entire
star forming regions, at long enough wavelengths to not be limited by
extinction, and determine the physical properties of whole populations
of young stellar objects (YSOs).  Trends in evolutionary state, mass, and spatial
distribution can help distinguish between star formation scenarios
such as competitive accretion \citep[e.g.][]{bonellbate} and
predestination \citep[in which cores determine stellar masses,
e.g.][]{tanmckee}.

One particularly vexing aspect of the complex and important topic of
star formation is that of triggering.  The formation of massive stars
feeds energy back into the nearby interstellar medium,
compressing, irradiating, and heating the natal molecular cloud.  This
feedback can have destructive or constructive effects, but it is not
clear which dominates in a given cloud or overall in a galaxy.
Although many examples exist of regions in our Galaxy and other
galaxies with suggestive spatial distributions of young and somewhat
younger stars, determining the relevant physical conditions and
timescales to make a strong argument for triggering is more
difficult. For some examples of these arguments, see the discussion of
the Galactic region W3/4 by \citet{oey}, 30~Doradus by \citet{walborn},
various regions and theory by \citet{deharveng}, the particular case
of cloud-crushing described by \citet{lefloch}, and further
theoretical considerations in \citet{elmegreen}.  This issue must be
clarified by making detailed assessment of whether triggered star
formation is happening in particular molecular clouds.

The Eagle Nebula (M16, containing the optical cluster NGC~6611)
contains a well-studied region of possible triggered star formation,
in the widely publicized ``pillars of creation'' or ``elephant
trunks'' imaged with HST by \citet[][hereafter H96]{hester96}.
NGC~6611 is a young open cluster, likely still in its formation
stages.  \citet{hillenbrand93} and \citet{dewinter97} found a spread
in ages of the optically visible cluster members: The massive
(M$_\star \gtrsim 10$\msun) stellar population is
$\sim$2$\pm$1~Myr old, with the most massive members apparently having
begun to evolve off of the ZAMS, and evidence of at least one evolved
(6~Myr old) 30\msun\ star \citep{hillenbrand93}.  There are also
optically visible (no longer embedded) intermediate mass stars
(3--8\msun) that are still evolving onto the ZAMS.  The famous
pillars are dense knots of molecular gas \citep{pound98} and dust
which shield the region behind them from the destructive radiation of
the O stars in the central cluster.  H96 identified ``Evaporating
Gaseous Globules'' (EGGs) in and around the pillars, and a central
question has been whether the EGGs are being compressed and forming
stars fast enough before they are photo-destroyed.  Detailed
near-infrared (NIR) searches find that about 15\% of the EGGs have
evidence of YSOs \citep{ma02,thompson02,sugitani02}.  Millimeter
continuum and molecular lines \citep{andersen04,white99,pound98}
reveal only one or two possibly younger objects in the pillars and at
their base.

Mid IR imaging offers new insight into this region in at
least two major ways.  First, not all objects with protostellar dust
show near-infrared (JHK) excess \citep[e.g. only half of YSOs with
10\um excess in Taurus-Auriga have NIR excess;][]{kenyon95}, and
modeling protostellar spectral energy distributions (SEDs) is much
more reliable with longer wavelength data
\citep[e.g.][]{whitney3,allen}. Second, the unprecedented mapping
speed of {\it Spitzer} allows imaging of a much larger area than the
previous NICMOS \citep{thompson02} and ground-based
\citep{sugitani02,ma02} studies centered on the pillars, permitting an
investigation of the protostellar population in the entire region.
This comprehensive assessment reveals star formation throughout the
region, with the least evolved objects most highly clustered together.
We also use mid-IR data to map the dust temperature and average
extinction; the YSOs are preferentially located in regions of
elevated dust temperature, and aside from NGC6611 which has cleared
its natal molecular material, also preferentially located in regions
of high extinction.

In section \ref{obs} we describe the observations and compare several
methods of selecting candidate YSOs. We will use the term YSO for any
evolutionary state -- heavily embedded, envelope dominated, disk
dominated, or remnant disk (commonly called Class 0,I,II, and III for
solar-mass objects).  Sections \ref{distro} and \ref{individual}
respectively describe the nature of the population and of individual
YSOs in the region.  The conclusions are summarized in
\S\ref{conclusions}.

%%%%%%%%%%%%%%%%%%%%%%%%%%%%%%%%%%%%%%%%%%%%%%%%%%%%%%%%%%%%%%%%%%%
%%%%%%%%%%%%%%%%%%%%%%%%%%%%%%%%%%%%%%%%%%%%%%%%%%%%%%%%%%%%%%%%%%%

\section{Observations and Analysis}
\label{obs}

M16 was observed as part of the Galactic Legacy Infrared Mid-Plane
Survey Extraordinaire (GLIMPSE), a {\it Spitzer} Legacy program
\citep{benjamin}.  The region was imaged with IRAC \citep{fazio} at
3.6, 4.5, 5.8 and 8.0\um with two 1.2s exposures at each location.
The point source sensitivities are approximately 0.5, 0.5, 1.0, and
4.0 mJy in the four filters, somewhat brighter in regions of intense
diffuse emission.  Sources were extracted from the images using the
GLIMPSE pipeline, built around a modified version of the PSF-fitting
program DAOPHOT \citep[][Babler et al in preparation]{stetson}.
Source detection and extraction is done in an iterative fashion.
Initial sources were detected at a 3-sigma level above the local
background.  These sources were subtracted from the image and then
a second round of detection was run at a 5-sigma level to reveal
sources that may have been missed in the wings of the PSF of bright
sources.  After the second round of sources were extracted a third
iteration of photometry was done to improve the fluxes.  In regions
of complex diffuse background, the source extractor can sometimes over-estimate
or under-estimate the point source flux (due to the difficulty of determining
the local background). To remedy this situation, aperture photometry was
done at the positions of extracted sources on the residual images (the
source subtracted images) to check for excessive over- or under-subtraction
of the source.  If the initial
extraction was indeed over- or-under subtracted, then the aperture
photometry was used to adjust the initial extraction value.
The pipeline has been rigorously tested against other
extraction software and using a network of calibrator stars with flux
densities predicted by spectral type and multiwavelength photometry
(M. Cohen, private communication).  Our simulations indicate that the
flux densities of truly pointlike sources are in general accurate to
$\sim$10\% even in regions of high diffuse background.  Zero point
flux densities (Vega magnitude system) used in the GLIMPSE v1.0
processing and this work are 280.9,179.7,115.0, and 64.13Jy for the
IRAC bands, and we used a zero point magnitude of 7.16Jy for our MIPS
24$\mu$m photometry \citep[The IRAC calibration of ][ is slightly different, and users of version 2 or higher GLIMPSE products should refer to the accompanying documentation]{reachcal}.  Sources detected at several wavelengths are
cross-correlated (bandmerged) using the SSC banderger software, which
is based on the 2MASS bandmerger.  Potential matches are considered
within 4'', and in confused regions, the association between all bands
which minimizes the positional chi-squared is chosen.

For this study, some of the sources of particular
interest (e.g. in the tips of the pillars) were re-photometered by
hand for this study, using irregular apertures and careful sky
selection in regions of irregular diffuse emission.  The results of
the manual photometry were for the most part consistent with the
pipeline extractions, but in a few cases of slightly extended sources,
the manual fluxes were substituted.  
It should be noted that photometry is extremely difficult
in very crowded regions with structured diffuse emission such as M16,
and although the GLIMPSE pipeline performs as well or better than any
other extractor we know of for IRAC data, one still needs to be very
careful about cross-correlating sources between filters and resolving
close pairs of sources at one wavelength but not another.  As we discuss below,
examination of each source's spectral energy distribution, and
comparing it to model SEDs, is actually a very good way to weed out
bad matches and questionably photometry.  Sources were bandmerged with
the 2MASS all-sky catalog \citep{2mass} using the GLIMPSE pipeline.

The GLIMPSE pipeline extracted over 42000 sources in the region
(within 0.45 degree of $l$=17.0, $b$=0.8, near NGC~6611), and we wish
to classify all of them in a consistent manner, as potential
YSOs, or reddened main sequence or post-main-sequence stars.  We
considered and compared four methods:
1) NIR-excess sources in a JHK color-color plot
2) sources with a 2--24\um spectral index greater than -2.0
3) sources in the IRAC color ranges of \citet{allen}
4) sources better fit by a protostellar than stellar model SED.

\begin{figure}
\rplotone{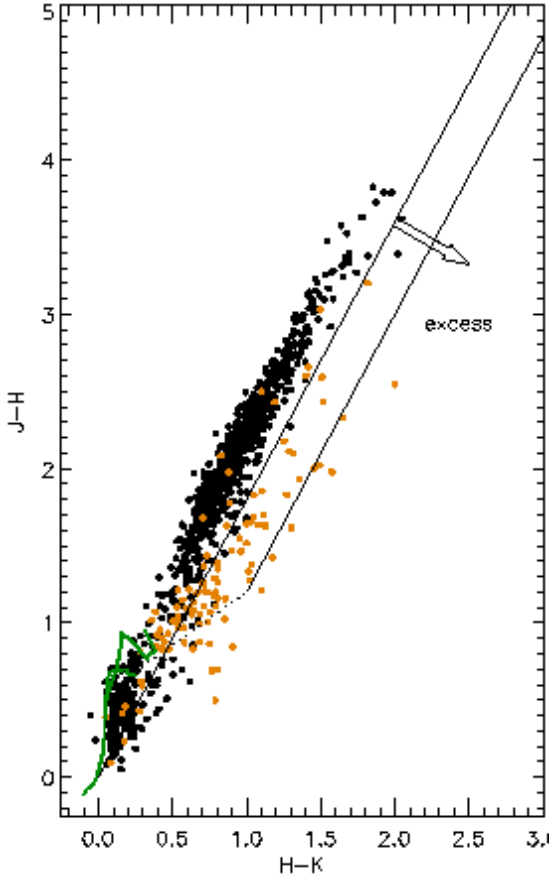}%m16mci.pstar.nir.4.eps}
\caption{\label{niryso} NIR color-color diagram with two lines
parallel to the reddening vector.  These separate the loci of dwarfs
and giants, T-Tauri candidates, and protostar candidates (the least
evolved young stellar objects, Class 0/I for solar-mass YSOs), from
top to bottom on the plot \citep[see text and][for
discussion]{meyer97}.  Candidate YSOs selected by SED fitting are
marked in orange (here and throughout, we will use the term ``YSO'' to include all evolutionary stages, not just the youngest). As noted in \citet{paper1} and shown observationally
\citep[e.g.][]{maercker,strom89}, only the small fraction of YSOs with
unobscured very hot dust have significant JHK excess.}
\end{figure}

Figure~\ref{niryso} shows a NIR color-color diagram dissected into
three regions in the usual manner.  The two dividing lines are
parallel to the reddening vector. We use E(J-H)/E(H-K)=1.75, as in
\citet{remylaw} and \citet{riekelaw}, although a range between 1.6 and
2.0 has been observed in different molecular clouds and the diffuse
ISM \citep{ccm,fitz}.  The line originating at (0,0) 
separates reddened main-sequence and giant stars
from T-Tauri candidates.  The other line separates T-Tauri candidates
from less evolved YSO candidates \citep[a dotted line marks the
classical T-Tauri locus, or experimentally determined color locus of unredenned T-Tauri stars -- disk-dominated YSOs are expected to lie near this line or along a redenning vector originating on the line][]{kenyon95,meyer97}.  We
define the ``NIR excess'' as the distance down and to the right of the
first line (reddened MS and giant stars have excess $<$ 0, YSO
candidates have excess $>$ 0.3):
\begin{eqnarray}
\mbox{excess} &=& {(H-K)\epsilon+0.59-(J-H)}\over{\sqrt{1+\epsilon^2}}, \\
\epsilon &=& E(J-H)/E(H-K)
\end{eqnarray}

Most other classification schemes for young stellar objects use
mid-infrared colors, or the mid-infrared spectral index.  Classically,
the 2-24\um spectral index (power-law exponent fitting
$\lambda$F$_\lambda \propto \lambda^\alpha$) has been used to classify
YSOs, and to define the ``Class I-II-III'' system (Lada, 1987).  
In the {\it Spitzer} era, many star formation regions have been
mapped with arcsecond spatial resolution using the 3.6--8.0\um IRAC camera
\citep{fazio}, so it is useful to consider methods of selecting
YSOs based in IRAC photometry alone.  However, Whitney et al.
(2003a,b,2004b) showed that the narrower wavelength range (3.6--8.0\um
v. 2--24\um) is more susceptible to 2-D effects (inclination), and
stellar temperature.  The 8.0\um band is also affected by the broad
10\um silicate feature that could be either in emission or
absorption.  More recently, we have computed a large grid of YSO
models and found that while the 3.6--8.0\um spectral index does correlate
roughly with evolutionary state (measured by circumstellar dust mass
and accretion rate), there is significant spread
\citep[][ Figure~11]{paper1}.  

\begin{figure}[h]
\rplotone{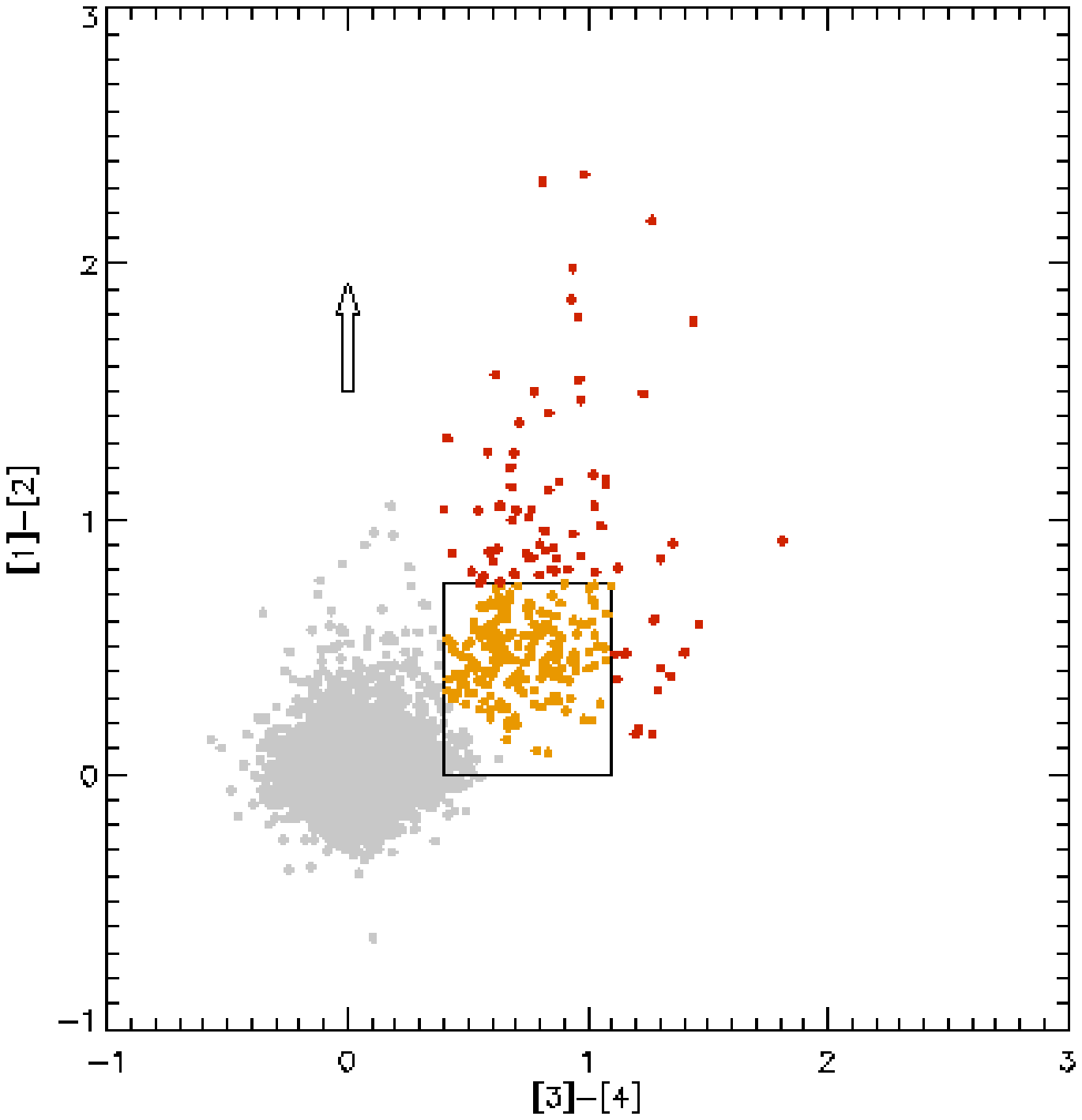}%m16mci.pstar.megeath.4.eps}
\rplotone{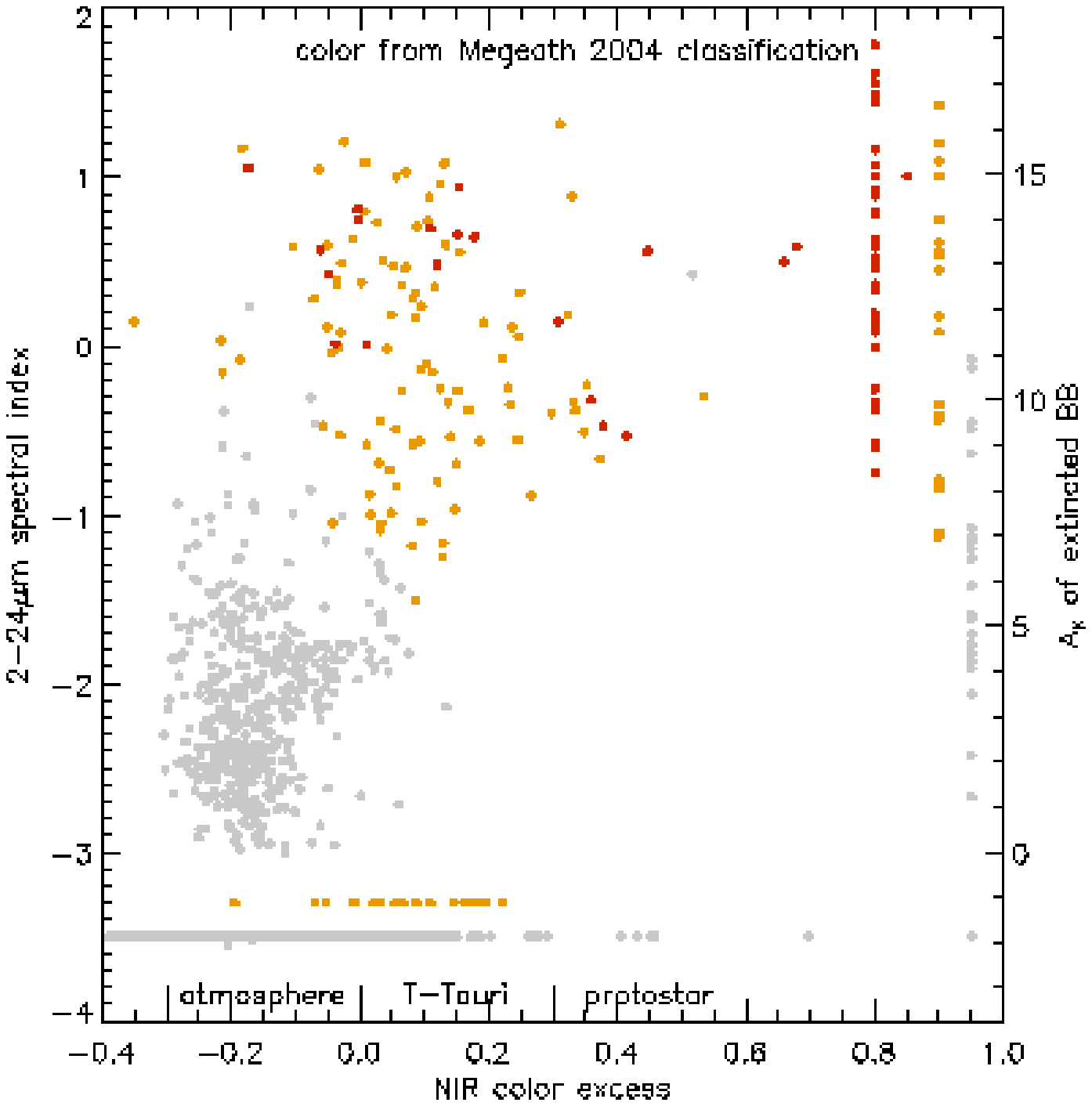}%m16mci.pstar.indexmegeath.4.eps}
\caption{\label{megeath} The first panel shows the IRAC color-selection
areas defined in A04,as applied to sources in the M16 region.  The
central box encloses candidate Class~II sources (orange),
and the redder sources candidate Class~I.  The grey points are
assumed to be stars with no circumstellar dust.  An extinction vector 
for A$_V$=30 using the measurements of \citet{remylaw} is shown.
The second plot shows
the A04 classification compared to the NIR color excess and MIR
spectral index.  Sources at the right edge of the plot do not have
J-band detections so no NIR excess could be calculated. 
Sources at the bottom do not have measurements at 2 or 24$\mu$m, so that particular spectral index cannot be calculated. The right-hand
axis shows the spectral index that a reddened Rayleigh-Jeans
photosphere would have at a given $A_K$ \citep[][measured $A_V=3.2$ towards NGC6611]{hillenbrand93}. }
\end{figure}

Classification based on IRAC colors uses more information about the
SED than a simple spectral index over the same wavelength range, and
may be more robust.  Figure~\ref{megeath} shows the regions defined
by \citet[][hereafter A04]{allen} \citep[see also][]{megeath1} in IRAC
color space.  Based on the solar-mass protostellar models of
\citet{calvet}, those authors suggest that the sources in the
central box are likely to be Class~II YSOs, and redder sources 
Class~I or younger.  
Our models do not show as neat a separation 
in color-color space as A04 due to the larger  
range of parameter space explored in our grid \citep{paper1}.  
The deeply embedded sources show color variations due to  
inclination effects from scattering in the bipolar cavities, and  
sources of all evolutionary stages show color variations due to inner  
hole size and stellar temperature \citep{whitney3}.
In Figure~\ref{megeath}, we compare A04
IRAC color-selected protostellar candidates with the spectral index
and the NIR excess (for sources with J,H, and K$_s$ detections).  The
IRAC-color classification is very consistent with having large
2-24\um spectral index, and somewhat consistent with having a NIR
excess, although the latter is not expected for all young sources.

In this study we use a different type of classification of YSO
candidates -- model SED fitting \citep{paper2}.  This is accomplished using a program
that fits all available fluxes to a large grid of model SEDs
\citep[][here we use fluxes between 1.25
and 24$\;\mu$m, but any between the UV and millimeter are usable]{paper2}.  
Our model grid consists of 2$\times$10$^6$ spectral
energy distributions emitted by 2$\times$10$^5$ young stellar objects
(YSOs; 10 inclination angles each).  The model grid is designed to
span both theoretically and observationally reasonable ranges of mass,
temperature, evolutionary state, disk mass, disk accretion rate,
envelope mass, and envelope accretion rate.  The colors of the model
grid also span the full range of observed colors of stellar and young
stellar objects.  
In
addition to the 200000 YSO models, our grid also includes 7853 stellar
model atmospheres \citep{brott} and 75 AGB star templates
derived from ISO spectroscopy and spanning the range of evolutionary
state and mass loss rate.  The fitter returns a best-fit for each
source as well as the distribution of templates and models that fit
with smaller than a user-selected value of $\chi^2$. 
We first select all sources that can be fit well ($\chi^2$ per
datapoint, or $\chi_\nu^2<2.5$) by an extinguished stellar atmosphere, and
consider these consistent with stars.  Of the remaining sources, if
one can be fit with $\chi_\nu^2<2.5$ by any YSO models, then we
consider it a candidate YSO.  If it cannot be fit well by either YSO
or stellar atmosphere, then it is likely poor photometry or
bandmerging.  This was confirmed by manual examination of the SEDs,
and the process is robust to changes of the threshold goodness-of-fit between
2.0$<\chi_\nu^2<$4.0 (the number of sources in each category
only changes by a few percent).
For this analysis we consider YSO models with distances between 1.5
and 2.5~kpc -- as discussed in \citet{paper2}, if the distance to the
sources can be constrained, the models can be better constrained by
the data.  
%We found two sources that were consistent with closer
%debris disks - they were poorly fit by YSO models at the distance of
%M16, but well-fit by a debris disk model at $\sim$0.5kpc.  We simply
%eliminated these from our analysis of the region.

Analyzing as much of the SED as possible has several advantages.
In crowded regions with bright complex diffuse emission such as M16,
examination of the SED provides an extra check to remove bad data
points.  For example, we found several sources that were resolved
from a close companion at H band and by IRAC, but not in the 2MASS
K$_s$ data, resulting in anomalously high K$_s$ fluxes (which would be
interpreted as a NIR excess).  SED analysis is robust to instrumental
artifacts and bad data in a single filter - color-color analysis
requires good photometry in four filters (the four IRAC filters for
example), whereas we can admit measurements with larger uncertainties
in some filters, but are protected against introducing many false
YSOs because if the resulting SED is inconsistent with any
stellar or protostellar model, it can be eliminated from the analysis.

Perhaps most interesting is the ability to understand the
constraints afforded by the data in as objective a way as possible.
If the photometry of a given source is consistent with a wide range of
circumstellar dust models, but inconsistent with any stellar
atmospheres, the source can still be classified as a probable
pre-main-sequence object, but at the same time it is known that the
data do not allow the evolutionary state (disk-only or accreting
envelope) to be determined.  This is philosophically similar to
choosing all sources with spectral index greater than some threshold,
or all sources with some threshold infrared excess.
In this manner, we use the model fitter to determine {\it upper
limits} to the evolutionary state of sources in M16: Namely, if a
source is not fit by any stellar atmosphere with $\chi_\nu^2<$2.5, 
but is fit at least that well by some YSO models, the
source is considered a YSO candidate.  If the source is fit by models
with accreting envelopes, but cannot be well-fit by any disk-only
model (no accreting envelope), then we consider the source a
candidate to be quite young (perhaps Class~I for a solar-mass object).
Finally, if the source can be fit by models with circumstellar disks
more massive than 10$^{-7}\times$M$_\star$ (where M$_\star$ is the
central source mass in the model), but is not fit by any models with
less massive disks, then we consider the source to be a candidate
accretion disk source (similar to Class~II for a solar-mass object).

\begin{figure}[h]
\rplotone{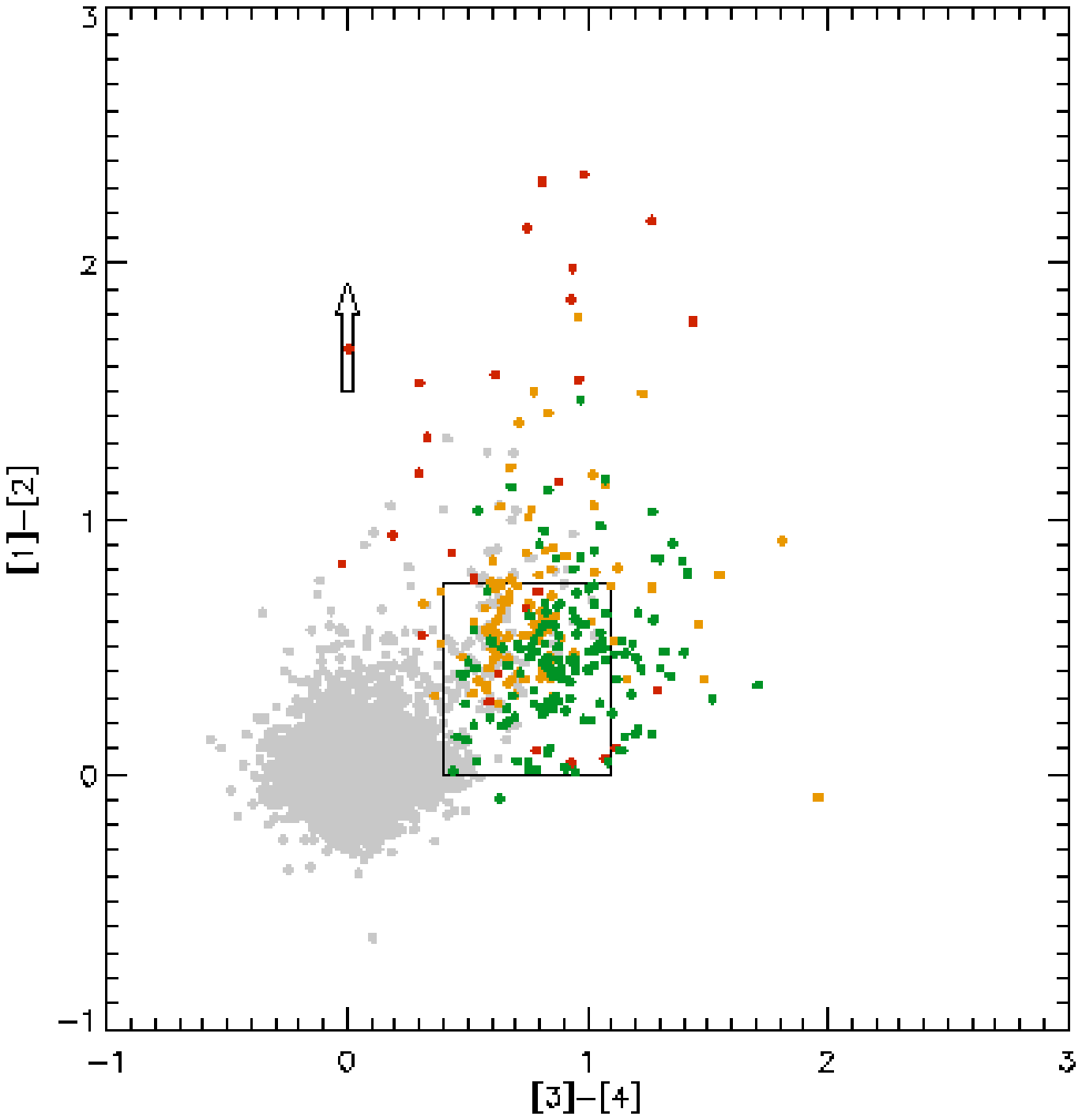}%m16mci.pstar.megeathfitter.4.eps}
\rplotone{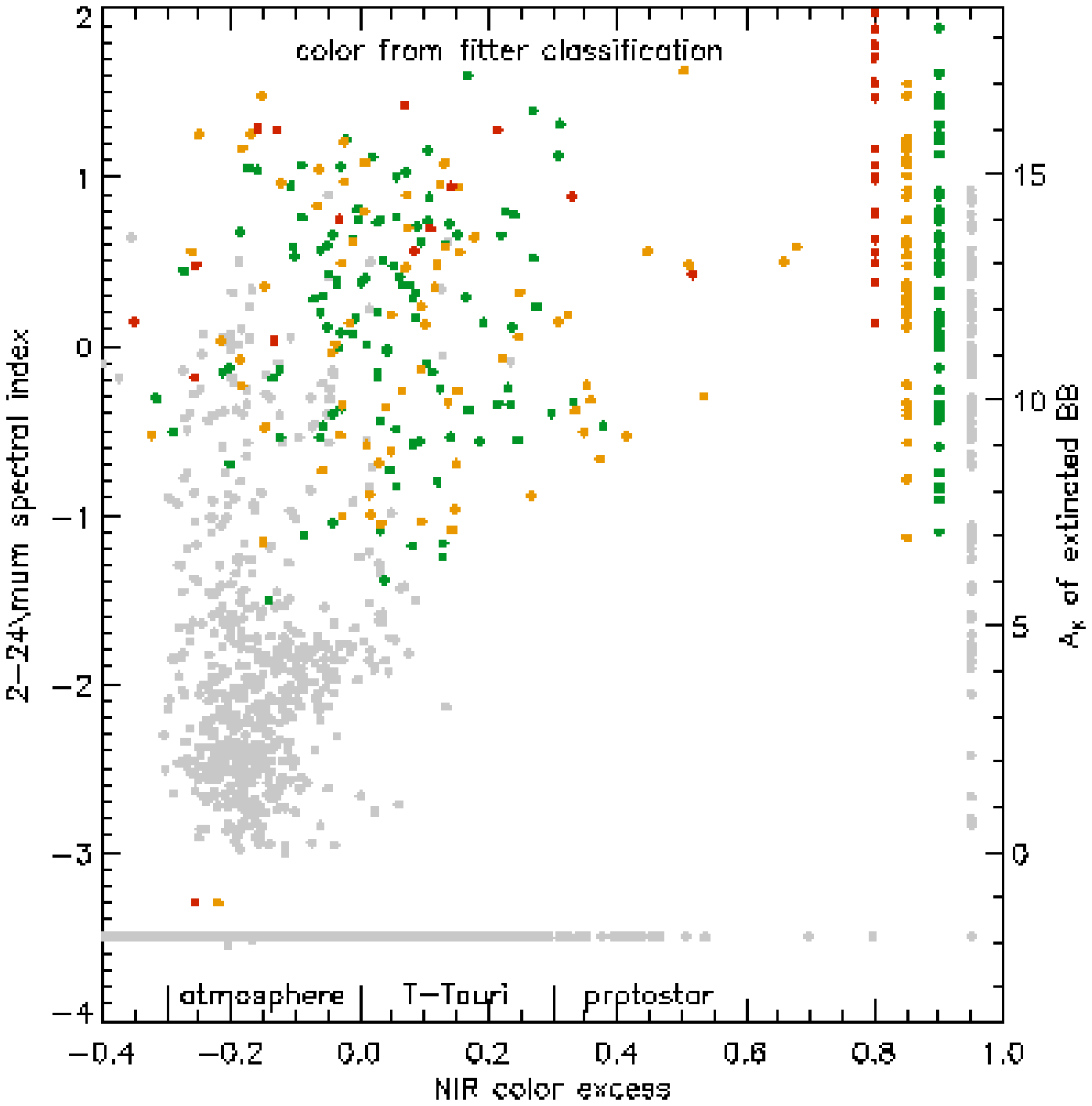}%m16mci.pstar.indexfitter.4.eps}
\caption{\label{indexfitter} Comparison of classification using our
model fitter with the color regions of A04 (first panel), and MIR
spectral index and NIR excess (second panel). Axes and annotations are
identical to Figure~\ref{megeath}.  An extinction vector for
A$_V$=30 using the measurements of \citet{remylaw} is shown on the
first panel.
Red points have a high probability of being relatively unevolved
with massive envelopes, orange points likely have still-accreting
circumstellar disks, and green points are YSO candidates for which the
existing data cannot place strong constraints on the circumstellar
dust distribution, but which are better-fit by YSO models than stellar
atmospheres (see text for details).  The sources may have more massive
envelopes or disks than shown, but the existing data do not allow that
to be unambiguously determined.}
\end{figure}

Figure~\ref{indexfitter} compares the classification suggested by our
model fitter with the NIR excess (for sources which JHK detection
allows calculation) and MIR spectral index.  Red points have a high
probability of being relatively unevolved with massive envelopes,
orange points likely have still-accreting circumstellar disks, and
green points are YSO candidates for which the existing data cannot
place strong constraints on the circumstellar dust distribution.  We
see that YSO candidates generally fall in the regions defined by A04.
The constraints suggested by the model fitter are less
well-defined than the box drawn by A04, as expected given a range of
inclination, stellar masses, and inner disk holes.  One should also
remember that the colors plotted are {\it upper limits} to the
evolutionary state, and the source may have more massive envelopes or
disks, but the existing data do not allow that to be unambiguously
determined. Foreground extinction also moves sources vertically in
this color space \citep[e.g.][]{remylaw}, which explains some
orange points lying above the box.  Some of the orange points to the right 
of the box (large [5.8]-[8.0], moderate [3.6]-[4.5]) have inner holes in their 
dust envelopes.
In the case of sources for which the data
can constrain the circumstellar dust distribution, there is a
correlation between less evolved objects (red points) and higher MIR
spectral indices.

Several characteristics of the fitter classification compared to a
classification based only on IRAC color are particularly important.
Comparing the second panels of Figures~\ref{megeath} and
\ref{indexfitter}, we see that the multi-band fitter finds many more
young sources that did not have JHK detections (the dots on the
right-hand side of Figure~\ref{indexfitter}).  Our fitter can
characterize a source with measured fluxes in as few as four bands
between J and [24], whereas a NIR excess characterization clearly
requires JHK detections and an IRAC-color characterization requires
detections in all four IRAC bands.  The latter can be difficult in
regions of high diffuse emission, like a star formation region.  (A
similar effect to fitting an SED to all bands can probably be achieved by
classifying sources in multiple color-color planes; Megeath et
al. 2005, Allen et al. 2005).
Careful examination of the entire SED allows the removal of
confused sources or those with poor photometry, which is less obvious
in a color-color (NIR or IRAC) plot.  Finally, comparison of the SED
with as many different YSO models as possible provides a relatively
unbiased way to assess the {\it degree} to which the data can
constrain the modeled nature of the source, beyond assigning a most
likely model based on colors.

%%%%%%%%%%%%%%%%%%%%%%%%%%%%%%%%%%%%%%%%%%%%%%%%%%%%%%%%%%%%%%%%%%%
%%%%%%%%%%%%%%%%%%%%%%%%%%%%%%%%%%%%%%%%%%%%%%%%%%%%%%%%%%%%%%%%%%%

\begin{figure*}
\plotone{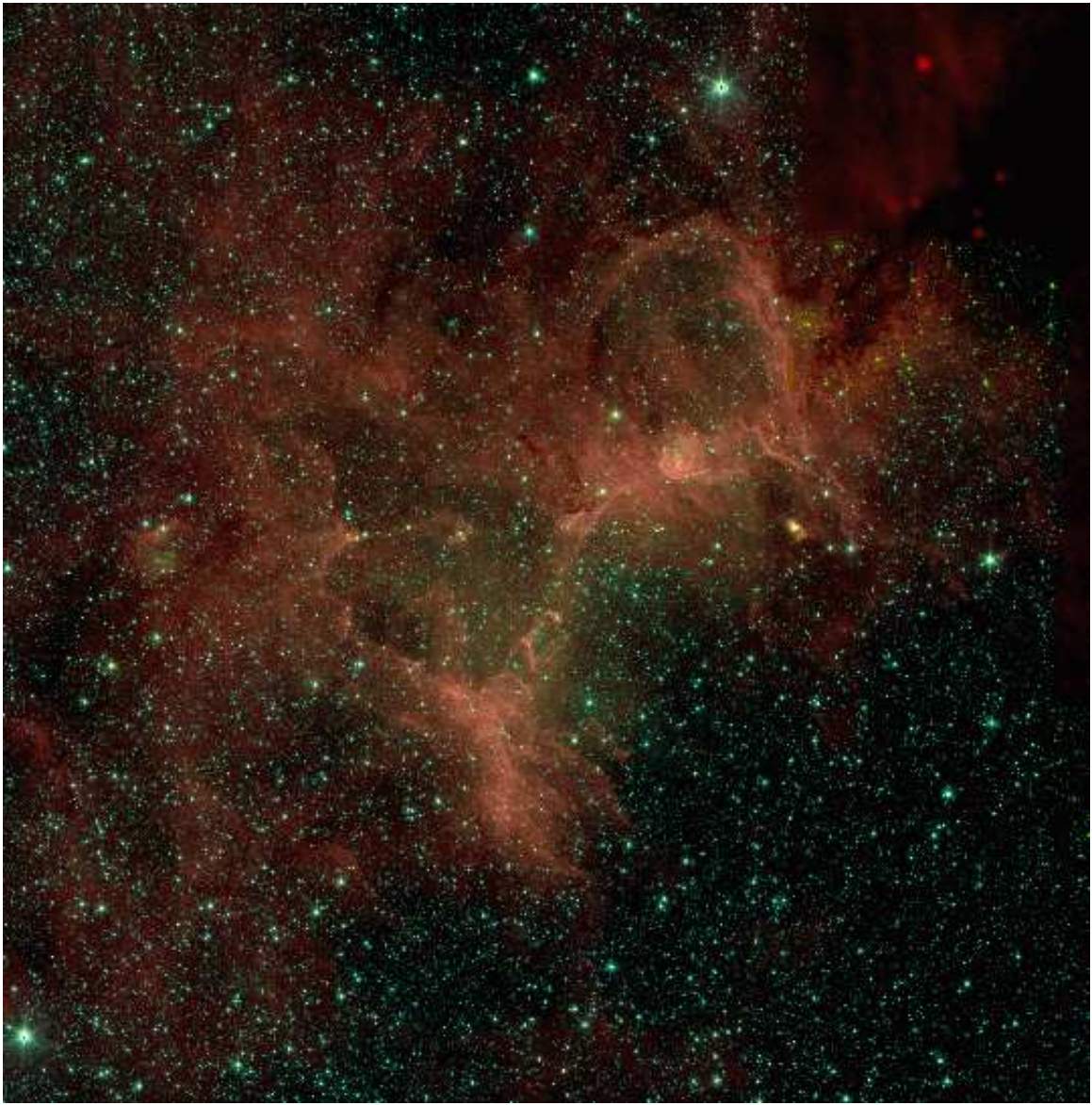}%glossy2b_75.ps}
\caption{\label{glossy}.  The Eagle Nebula as seen by IRAC (RGB =
[8.0],[4.5],[3.6]). IRAC [8.0] has been supplemented with MSX A (8$\;\mu$m) in the NW corner.}
\end{figure*}

%\clearpage
\section{The distribution of YSOs in the Eagle Nebula}
\label{distro}

\begin{figure*}
\plotone{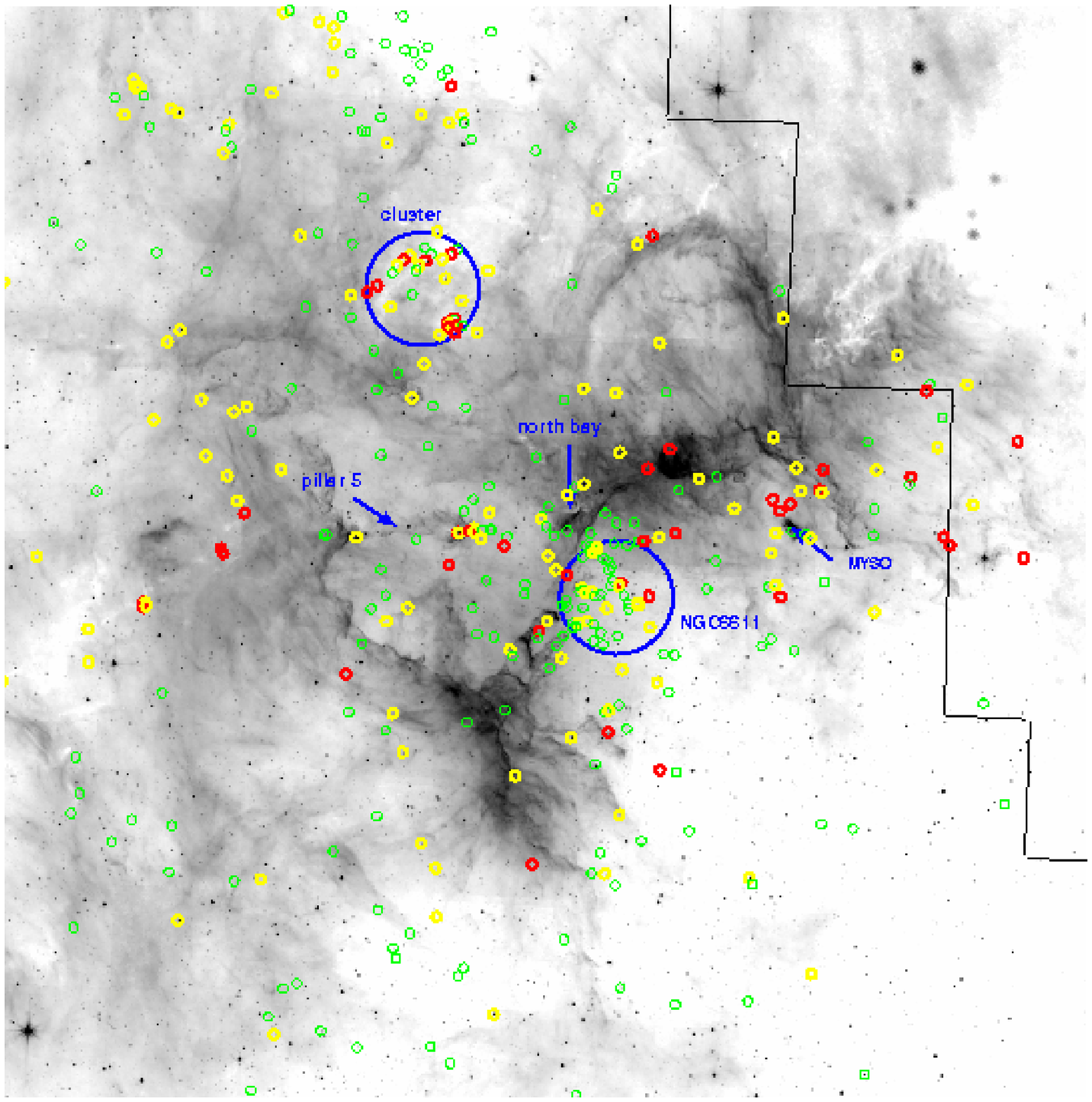}%ysoa_75.ps}% switch to 300 dpi version for submission.
\caption{\label{overview} IRAC [8.0] image of M16 with stars of
various inferred evolutionary states marked in different colors.  
Sources that can be constrained to have relatively massive
accretion disks are yellow, and those that in addition likely have
massive circumstellar envelopes are red.  YSO candidates for which the
data do not strongly constrain the mass of circumstellar dust (merely
that they have some, and are inconsistent with stellar atmospsheres)
are green.
NGC~6611 and a new cluster of YSO candidates to the north are
marked with large blue circles.  The edge of the region for which coverage
exists in all IRAC bands and with MIPS at 24\um is marked as a thin
black line.
}
\end{figure*}

Figure \ref{glossy} shows the Eagle Nebula in the [3.6], [4.5], and
[8.0] bands.  The optical open cluster is the concentration of blue
stars to the NW of the HST pillars (marked in Fig~\ref{overview}).
Protostellar candidates selected using our fitter are shown on a [8.0]
image in Figure~\ref{overview}.  Many of the protostellar candidates
are located in the NGC~6611 cluster, which is known to be very young.
In fact, nearly all of the sources in the NGC~6611 region are
classified as YSO candidates: 46 of the 47 sources with high
signal-to-noise detections in at least 4 bands between 1 and 10\um
and located in a 4\arcmin\ diameter region around NGC~6611 were
classified as YSO candidates.  This is consistent with a very high
L-band (3.5$\;\mu$m) excess fraction (85$\pm$5\%) determined by
\citet{oliviera} and the large number of emission-line stars seen by
\citet{dewinter97}, implying that circumstellar disks are fairly
robust, even in a cluster environment.

Our analysis reveals a new cluster to the NE (at 18h19m10s, -13d36m,
approximately 7\arcmin\ in diameter).  Located in a dark region
indicative of a dense cloud, the new cluster is likely younger than NGC~6611 -- 
as seen in Figure~\ref{overview}, most of the new cluster's sources can be constrained 
to have accreting disks (yellow) or envelopes (red), compared to NGC~6611
which has many sources consistent with remnant disks (green).
There is also a concentration at the head of a large pillar at
18h19m05s -13d45m20s (``Pillar 5''; we analyze this region in more
detail below, \S\ref{p5}).  Finally, there is a low density of
scattered protostellar candidates across the region shown. This
apparent interspersed ongoing star formation (also noted in the giant
HII region RCW49 by \citet{whitney_rcw49} and in Orion by
\citet{megeathorion}) could possibly be a superposition of background
objects, but at least some of them are likely YSOs, indicating
widespread low-level activity throughout the cloud.

%%%%%%%%%%%%%%%%%%%%%%%%%%%%%%%%%%%%%%%%%%%%%%%%%%%%%%%%%%%%%%%%%%%

\subsection{Spatial Relationship of YSOs to the Molecular Cloud}

\begin{figure}
\plotone{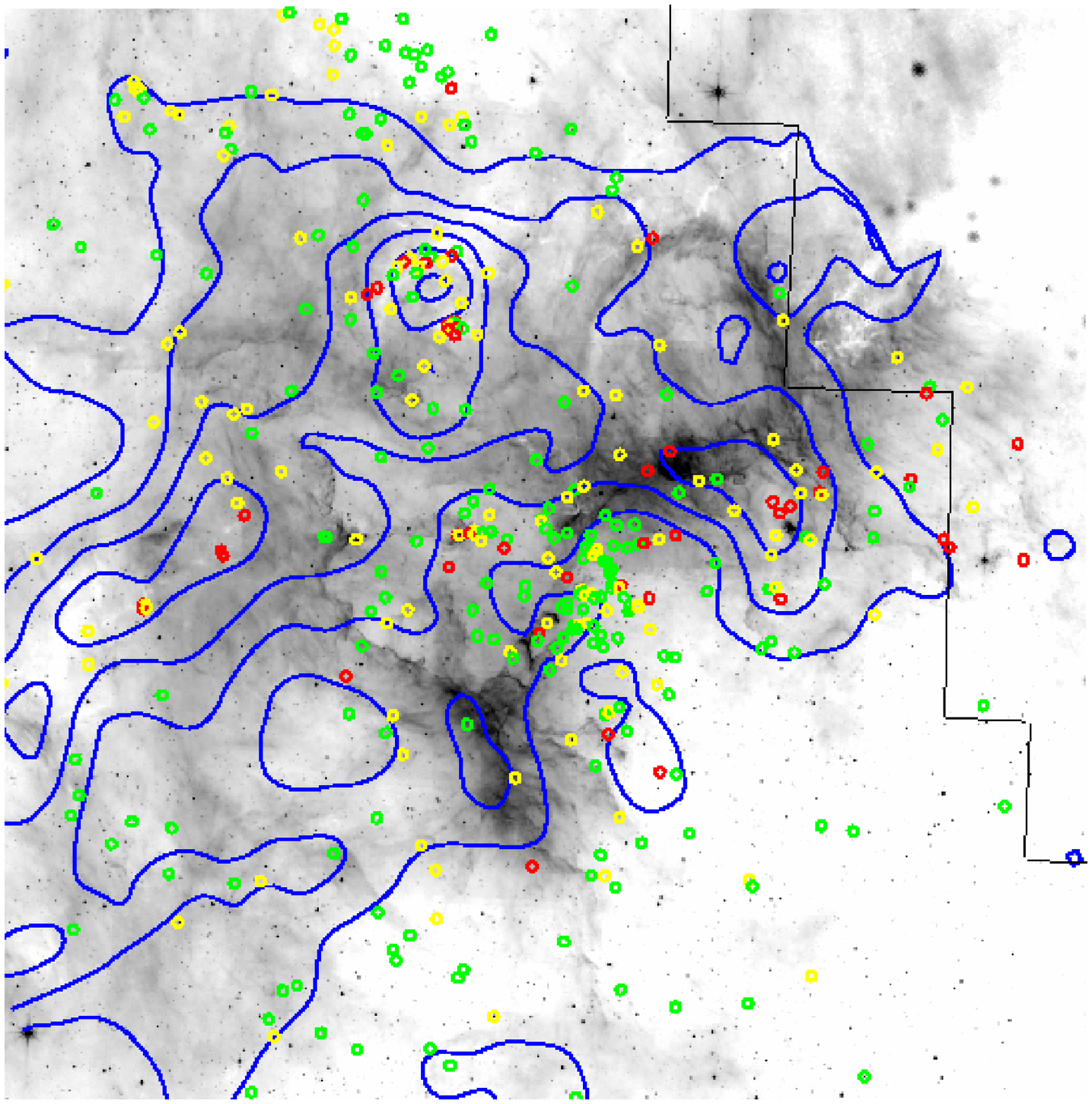}%ysoext_75.ps}
\caption{\label{extinction} [8.0]\um image of M16 with contours of
average extinction and the location of protostellar candidates (color
as Fig~\ref{overview}).  Contour levels are A$_V$=10.5, 12.5,
14.5, 16.5, 18.5, 20.5.  NGC~6611 is located in a low-extinction
region, likely evacuated by the cluster itself.  There is a
concentration of dense gas to the NE of NGC~6611, and more star
formation on the other (N) side of the cloud, in a region of somewhat
lower extinction. The dark area to the NW is missing IRAC data, so the
blue contours end artificially due to lack of point source photometry}
\end{figure}

A byproduct of our fitting program is the best-fit extinction to each
source, including the large number of extinguished main-sequence and
giant stars along the line-of-sight.  These values can be binned to
produce a map of the average extinction, which is shown as contours in
Figure~\ref{extinction}.  We expect that the molecular cloud
associated with star formation in M16 dominates the extinction
measured with this procedure, and indeed the features match other
indicators of dense material.  NGC~6611 is located in a low-extinction
region, likely evacuated by the cluster itself.  High-extinction
regions correspond reasonably well with the boundaries of the bubble
delineated by PAH emission in the the IRAC bands and scattered light
in the NIR.  
%In fact, the average extinction map highlights what is
%the apparent SW edge of the bubble more clearly than the PAH emission.
There is a concentration of dense gas to the NE of
NGC~6611, and more star formation on the other (north) side of the cloud, in
regions of somewhat lower extinction.  We compared the spatial
distributions of YSOs as a function of evolutionary state and
mass, and found no obvious variations with either parameter other than 
the likely younger NE cluster.  
Since our maximum extinction through the cloud is A$_V\simeq$20,
which is 1 magnitude of extinction at IRAC wavelengths
\citep{remylaw}, we can still detect YSOs down to masses of
about 0.5\msun, and we do not expect to be missing any significant
star formation in the region.

\begin{figure}
\plotone{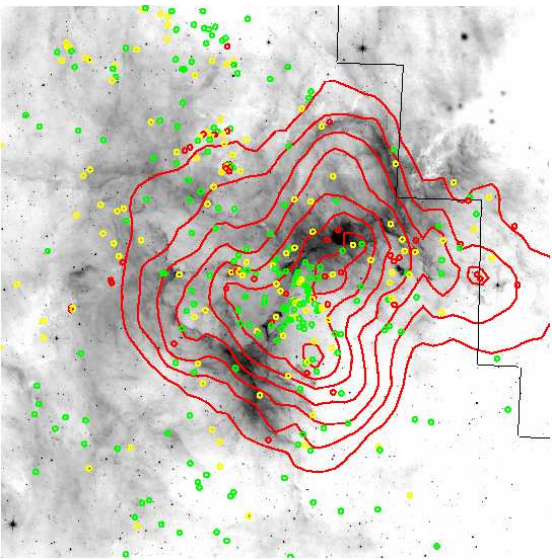}%ysotemp_75.ps}
\caption{\label{dusttemp} Contours of dust temperature calculated from IRAS
60\um and 100\um images, overlaid on the IRAC [8.0] image of M16.
Contours levels are 32, 33.3, 34.7, 36, 37.3, 38.7, and 40K.}
\end{figure}

The combination of mid-IR imaging at different wavelengths allows
one to directly compare the protostellar population to the properties of
the molecular cloud in which they are forming.  First,
we use IRAS 60\um and 100\um images to determine 
the temperature of the dust:
\[ T_{dust} = {{hc/k(1/\lambda_2-1/\lambda_1)}\over
{(3+\beta)\ln(\lambda_1/\lambda_2) + \ln(F_\nu(\lambda_1)/F_\nu(\lambda_2))}},\]
where $\beta$ is the emissivity index,
\[ F_\nu \propto \left[{{2hc}\over{\lambda^3(e^{hc/(\lambda kT_{dust})}-1)}}\right]\lambda^{-\beta} \]
\citep[see e.g.][]{li_iras,complete}.  Figure~\ref{dusttemp} shows
contours of the dust temperature relative to the locations of
YSOs and PAH emission.  Clearly, the warmest dust is heated by
NGC~6611, as one would expect.  It is also evident by the NE and SW
extensions of the contour shapes 
centered on the ``pillars of creation''
that the cavity presumably blown by
that young open cluster is warmer than the surrounding denser
molecular cloud. There is also some evidence of heating to the west of
NGC~6611, perhaps by embedded sources like the massive YSO that
we find there (\S\ref{massive}).
\label{tdust}

\begin{figure}
\plotone{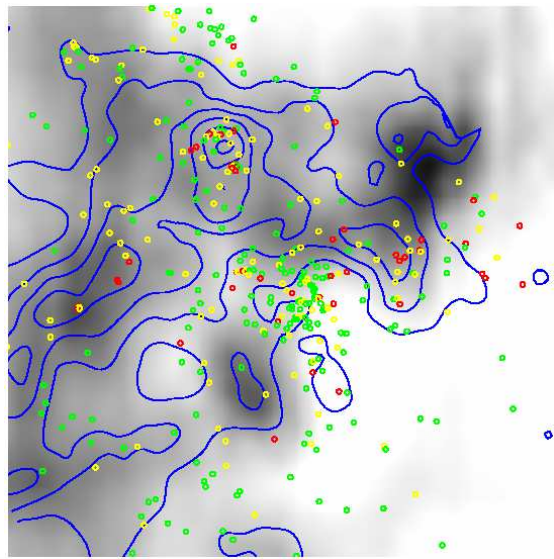}%ysotau_75.ps}
\caption{\label{tauext} Contours of extinction calculated by fitting
42000 stellar point source SEDs, overlaid on a gray scale image
of the optical depth of emitting dust calculated from IRAS 60\um and
100\um images.  Candidate YSOs identified by our fitter are
marked as circles colored as Fig~\ref{overview}. The dark area to the
NW is missing IRAC data, so the blue contours end artificially due to
lack of point source photometry. }
\end{figure}

The optical depth of a molecular cloud can also be calculated from the
dust emission, and this compares well with our map calculated using
embedded stars.  Given a temperature, the optical depth of the emitting dust
is simply $\tau_\nu = F_\nu / B_\nu(T_{dust})$, where
$B_\nu(T)$ is the Plank function.  We use the 100\um image to
calculate $\tau$, but the results using the 60\um are very similar.
We used $\beta=$2 for these calculations; choosing $\beta=$1 increases
the temperatures by 15\% and decreases the optical depths by a factor
of 2, but does not affect the {\it relative} optical depths, which
features we are most interested in examining here.
Figure~\ref{tauext} compares the optical depth calculated by the
fitter using point sources photometered by GLIMPSE to the optical
depth calculated from cool dust emission - most features agree well,
including the cavity and dense ridge.  The new NE cluster region shows
high extinction in the point sources, but low optical depth in the
dust.  

%%%%%%%%%%%%%%%%%%%%%%%%%%%%%%%%%%%%%%%%%%%%%%%%%%%%%%%%%%%%%%%%%%%

\subsection{Clustering}

\begin{figure}
\rplotone{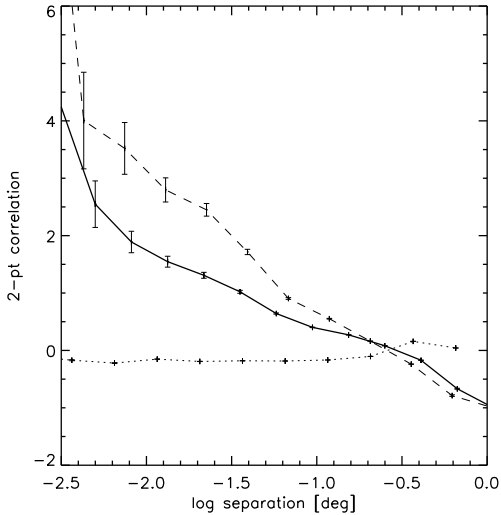}%m16.correl.eps}
\rplotone{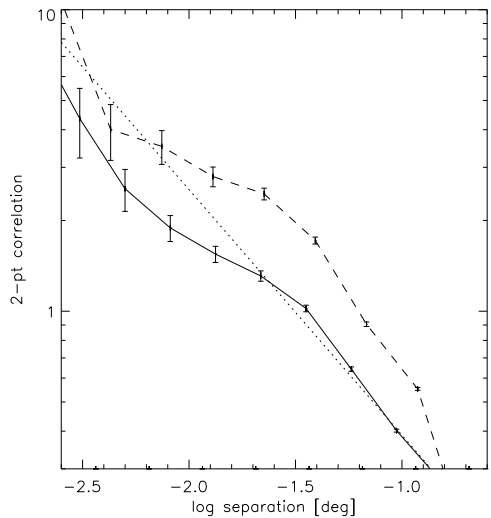}%m16.correl.log.eps}
\caption{\label{2pt} The two-point correlation function of YSO
candidates in M16.  The correlation function is plotted on a linear
scale in the first panel, and on a logarithmic scale in the second
panel.  The dashed line is the correlation function of only the less
evolved YSO candidates (best-fitting models have dust envelopes with
nonzero accretion rates). The dotted line on the first panel is the correlation 
function for all sources in the region, which are uncorrelated.
%A weak signature of a
%preferred scale is present at $\theta\sim$10$^{-1.3}$deg = 1.8pc (at
%d=2kpc).  
The dotted line in the second panel corresponds to a
power-law correlation function with index -0.8}
\end{figure}

Stars are generally thought to form in cluster environments, so
greater clustering in a population should be a sign of youth.  The
clustering properties of young stellar objects should also reflect any
preferred spatial scales associated with the formation mechanism, if
those exist.  In Figure~\ref{2pt} we show the two-point correlation
function of our YSO candidates in the M16 region.  We count the number
of pairs of YSOs as a function of angular separation $\theta$, binned
logarithmically -- i.e. $N(\theta)$ is the number of pairs with
separation between $\theta$ and $\theta + d\log\theta$.  This
distribution is compared to the distribution of separations of a
random field of stars with the same average stellar density
$N_r(\theta)$; the correlation function is $N/N_r-1$.  The maximum
value of the correlation function is 
%fairly high 
comparable to peak values of %\citep[$\sim$8,
%compared to e.g. 
0.5, 1.5, 6, and 15 in subclusters in the Rosette
Molecular Cloud \citep{rosette}, reflecting the strong degree of
clustering.  The correlation function is only roughly consistent with
a power-law (Fig~\ref{2pt}, second panel), which would mean there is
no preferred spatial scale for star formation in the region.  The
slope of the correlation function in logarithmic space (-0.8) is also 
comparable to other clusters \citep{rosette, scalo}, also
indicating fairly strong hierarchical clustering.  
Typical optical depths in region are
A$_V$=15--20, corresponding to N(H$_2$)$\sim$4$\times$10$^{20}$ in the
molecular material.  The cloud appears filamentary, with thickness
about 4pc, so if the depth along the line of sight is similar, the
average volume density in the cloud is n($H_2$)$\sim$150cm$^{-3}$,
with a thermal Jean's length of 2pc (R$_J\simeq$4.8pc(T$^3$/n)$^{0.5}$ if
T is expressed in K and n in cm$^{-3}$).  
At a distance of 2kpc, this corresponds to a size scale of
10$^{-1.3}$; there may be some signature of a turnover in the
correlation function at about 10$^{-1.4}$ degrees, but the result
is not definitive.  There are also structures in the region of about
that size (for example the bubble around NGC~6611, and star-forming
pillar~5, (\S\ref{indiv}) and the correlation function could be
reflecting those structures. 

\begin{figure}[h]
\rplotone{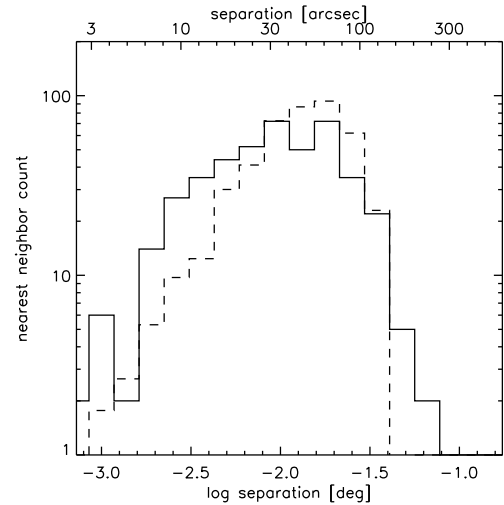}%m16.nearest.eps}
\caption{\label{nearest} The distribution of nearest neighbor
distances of YSO candidates in M16 (solid line) compared to the
distribution for a random star field (dotted line).  The YSOs show an
excess at separations $\sim$10\arcsec, or 0.1pc at a distance of 2kpc.
}
\end{figure}

The distribution of nearest neighbor distances of young stellar
objects can potentially reflect a fragmentation scale.  We show that
distribution in Figure~\ref{nearest}, compared to a random
distribution. The YSOs show an excess at separations $\sim$10\arcsec,
or 0.1pc at a distance of 2kpc.  This is consistent with the
fragmentation scale in the hot dense gas such as in the heads of the
pillars and molecular cores near HH216, which are warm and dense
\citep[T$\sim$40K, n$\sim$10$^{5}$cm$^{-3}$,][yielding
R$_J\sim$0.1pc]{pound98,andersen04}.
The signature of the parent molecular cloud's Jeans length has been seen in 
the separation of YSOs before, notably by \citet{tiex} in NGC2264.
The YSOs with nearest neighbor distances in the range 10$^{-2.7}$
to 10$^{-2.2}$ degrees are located in the densest filamentary parts of
the molecular cloud to the north (seen as dark lanes in the 8\um
image), associated with the HST pillars and pillar 5, and also in
NGC6611.  This association with dense molecular material suggests that
the separation and Jeans length may not merely be a coincidence.  It
is less clear why many stars in NGC6611 have that nearest neighbor
separation - perhaps conditions in the cloud at the time of formation
of NGC6611 were similar.

%\clearpage
\section{Specific Objects and Regions}
\label{individual}

%%%%%%%%%%%%%%%%%%%%%%%%%%%%%%%%%%%%%%%%%%%%%%%%%%%%%%%%%%%%%%%%%%%

\subsection{The HST ``Pillars of Creation''}

Using ISOCAM, \citet{pilbratt98} stated that the lack of mid-infrared
emission along the length of the pillars indicated that there was not
enough dense star-forming material present to form new stars.  The
resolution and sensitivity of the ISOCAM images was not sufficient to
rule out very embedded or low-mass stars (the sensitivity limit of
20mJy at 6.7\um in bright regions corresponds to a 4\msun\ YSO,
A$_V$=30, 10$^5<$age$<$10$^6$yr).  The higher resolution and
sensitivity of the IRAC images allows us to confirm Pilbratt's
statement with much more stringent upper limits - we detect no new
mid-infrared point sources down to a limit of $\sim$1mJy in the [3.6]
band, and $\sim$5mJy in the [8.0] band (limited by the strong diffuse
background emission), corresponding to 0.4\msun.

\begin{figure}
\plotone{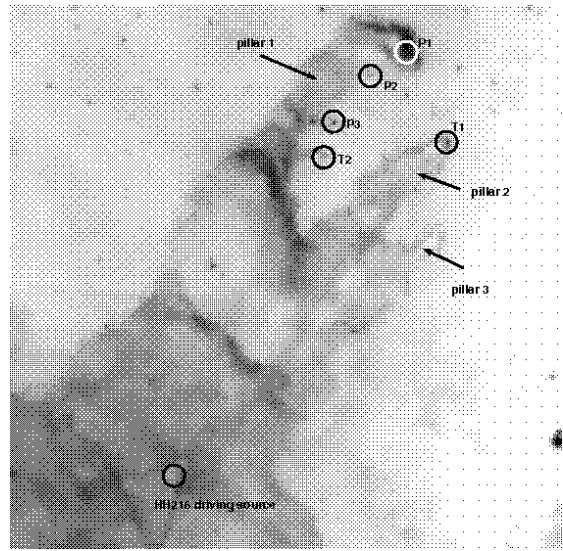}%trunk_75.ps}
\caption{\label{indiv}Individual sources in the HST pillars region, 
shown on IRAC [8.0] image.
See text for discussion of their nature, and Tables~\ref{ysotable1} and
\ref{ysotable2}
for coordinates and physical parameters.}
\end{figure}

There are a handful of YSOs in the pillars, already identified
in the NIR \citep[e.g.][]{sugitani02,andersen04,thompson02}.  Figure
~\ref{indiv} shows the locations and names of notable objects.  Our
new mid-IR flux densities for many of these allows a more precise
determination of their likely physical parameters than was possible
from NIR data alone.  We use the grid of YSO models described in
\S\ref{obs}, and demonstrate how to make a careful assessment of all
the models that are consistent with the data.  Particularly important
are any age constraints that could strengthen or weaken arguments
that the YSOs in the trunks are {\it triggered} by the action of
NGC~6611.  The evaporation timescale for the trunks is
2$\times$10$^7$yr, but internal velocity gradients suggest that they
will disperse or be torn apart in only 10$^5$yr \citep{pound98}.  The
trunks appear to be in rough pressure equilibrium with the surrounding
\ion{H}{2} region
\citep[P/k$\simeq$8$\times$10$^7$cm$^{-3}$K, ][]{pound98}, so there is
no well-defined ``crushing'' timescale.

\begin{figure}[h]
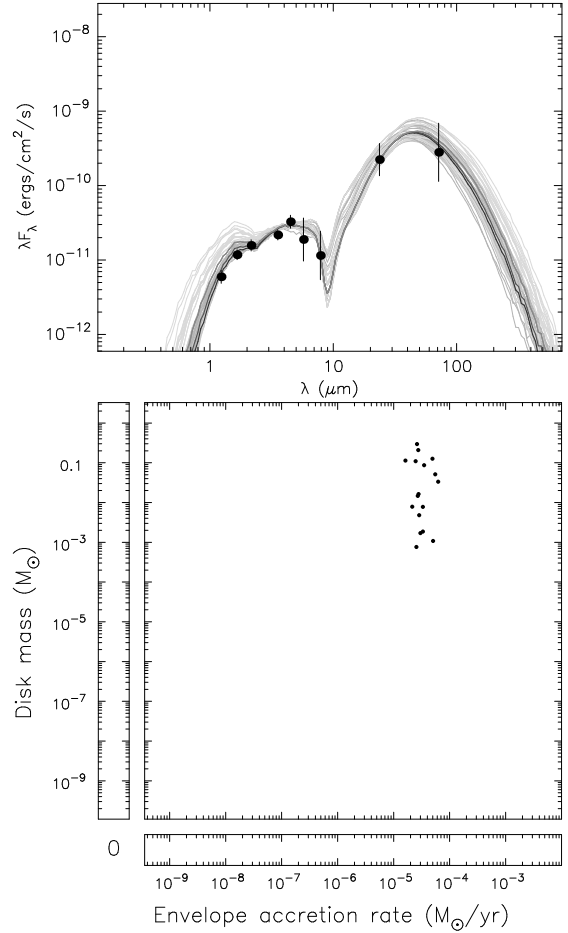

\rplotone{f12a.eps}%P1.eps}
\rplotone{f12b.eps}%P1.mm.eps}
\caption{\label{p1} Source P1.  The first panel
shows the data and the model SEDs that fit better than $\chi_\nu^2<3.7$.  
The second panel shows the envelope accretion rate \.{M} and disk mass 
M$_{disk}$ for those models.  
%The grey scale shows the density of models in the grid.  
The plot range is that of the entire model grid \citep[see][for
details]{paper1}.  The bars along the axes are for models with zero
\.{M} and M$_{disk}$, i.e. if there were well-fitting models with zero
disk mass they would be in the bar along the $x$ axis (this will be true
for some sources below).
Nearly all of the models that fit the data for this 
source well have high accretion rates and disk masses, consistent with 
the large spectral index and a relatively young evolutionary state.
}
\end{figure}

The most prominent young objects in the pillar region are P1 and T1 at
the tips of the first and second most northern pillars ($\Pi_1$ and
$\Pi_2$), respectively.  At our modest 1.5\arcsec\ spatial resolution,
the P1 source is the same as that identified in the NIR by
\citet{sugitani02}, and as M16ES-1 by \citet{thompson02}.
\citet{fukuda02} detect 2.7mm sources further
behind P1 in the pillar and suggest that these are distinct
YSOs, but our data does not support that suggestion.  
Figure~\ref{p1} shows 
the MIR SED and protostellar models consistent with the data.
The 2--24\um spectral index of the source is $\alpha$=1$\pm$0.2,
Class~I in the usual nomenclature \citep[][Class~I: 0$<\alpha<$3,
Class~II: -2$<\alpha<$0, Class~III: -3$<\alpha<$-2]{lada87}.  Source
P1's bolometric luminosity (estimated by $\Sigma\nu F_\nu$) is
44$\pm$10\lsun\  (observed quantities are tabulated in
Table~\ref{ysotable1}).  Using the pre-main-sequence evolutionary
tracks of \cite{siess}, this is consistent with a 2.5--4\msun\ 3\e{5}
old YSO (the tracks are quite close together so a small
uncertainty in age or luminosity leads to a large uncertainty in
mass).
The YSO model fitter produces similar results, as tabulated in
Table~\ref{ysotable2}.  Models with masses between 1.6 and 6.4\msun\
fit the data with $\chi_\nu^2<$3.7, and the best-fitting model has a
mass of 4.5\msun (for this analysis we consider all models which fit
to $\chi_\nu^2<$3.7 because that represents a probability of 1/1000,
and we wish to be conservative in our constraints on physical
parameters).  Those models fit with a foreground (line-of-sight)
extinction A$_V\simeq$4.  They have envelope accretion rates \.{M}
between 1\e{-5} and 7\e{-4}\msun/yr, also consistent with a Class~I source.
\citet{thompson02} derive a stellar luminosity of 200L$_\sun$, and
comment that the lack of Pa$\alpha$ emission suggests a group of
several lower-mass objects.  Our fitted mass is that of a very late B
star, and low ionizing flux.  \citet{ma02} find a somewhat higher mass
(10M$_\sun$, a B1 star). Those authors also derive an extinction of
$A_V$=27, {\it if all the J and H band light is photospheric}.  The
extinction from our model fit of $A_V\simeq$4 listed in the Table is only
the foreground extinction to the YSO; the total extinction including
foreground and circumstellar dust is $A_V\simeq$40%\pm$10.

%\clearpage

%\rev{ I'm thinking of deleting P2, since we can't really constrain it very well}

Another red source, which we and \citet{sugitani02} call P2, is
located (at least in projection) below the tip of Pillar~1 with
respect to NGC~6611, and noted to have high extinction by \citet{ma02},
who call it E23 in correspondence with the H96 EGG of that
designation.
This source has a spectral index of 0.3$\pm$0.2 and a bolometric
luminosity of 5.3$\pm$2.8\lsun\ - the inability to determine a reliable
24\um flux makes it more difficult to constrain this object's
luminosity and evolutionary state.  Using the PMS tracks as for P1 and
assuming an age of a few 10$^5$ years suggests that this source is
about 1\msun.  Using only a 24\um upper limit does not allow the model
fitter to constrain the parameters very well either, with accretion
rates between zero and 10$^{-3}$ and masses between a few tenths and a
few solar masses fitting the data with $\chi_\nu^2<3.7$.  The mid-infrared
excess and positive spectral index suggest that accretion is still
continuing in this object (and no model without any accretion disk is
consistent with the data).  
\begin{figure}
\rplotone{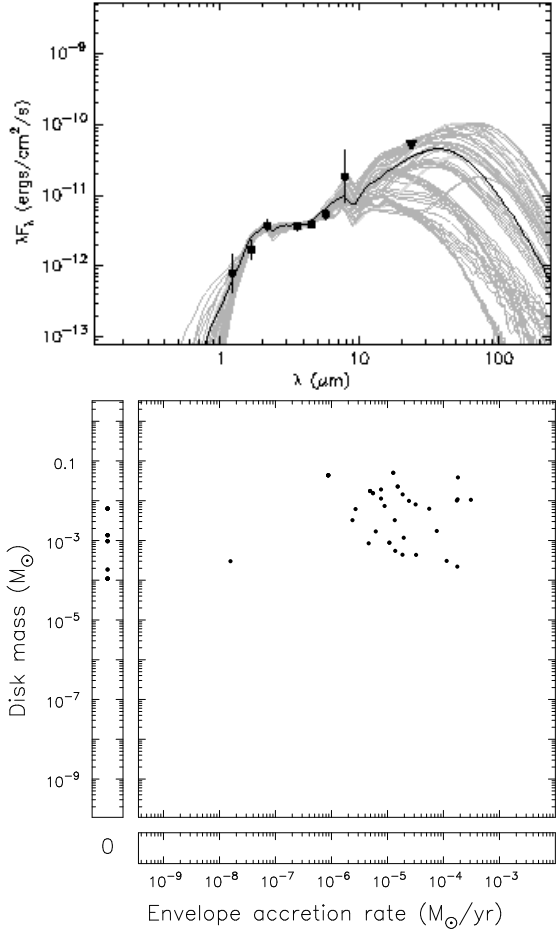}%P2.eps}
\rplotone{f13b.eps}%P2.mm.eps}
\caption{\label{p2} SED and model fits to the source P2. Plots and annotation
as Fig~\ref{p1}.}
\end{figure}

%%%%%%%%%% tables were here

\begin{figure}
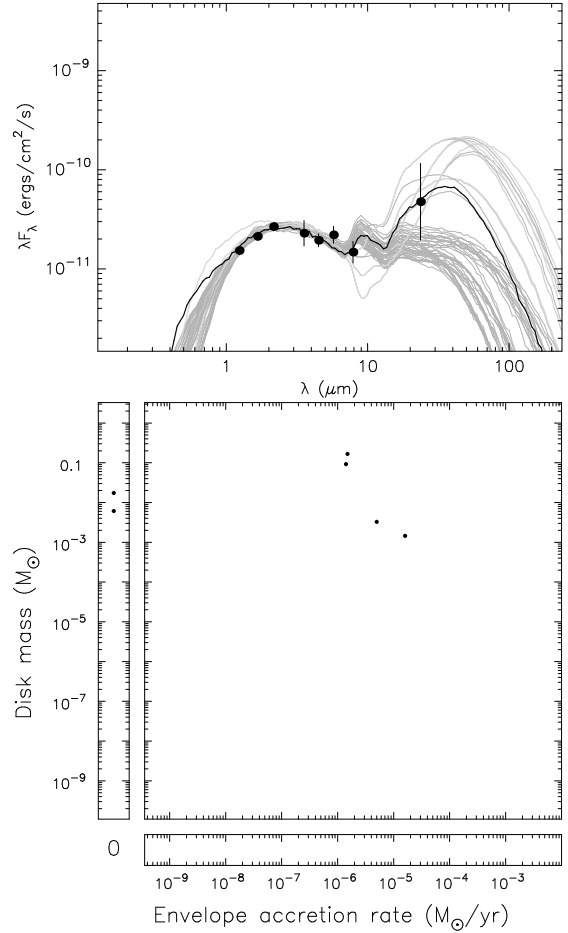

\rplotone{f14a.eps}%T1.eps}
\rplotone{f14b.eps}%T1.mm.eps}
\caption{\label{t1} Model fits to the source T1. Plots and annotation
as Fig~\ref{p1}.}
\end{figure}

The source at the tip of the central pillar (T1) was classified as a
somewhat older source, a candidate T-Tauri star, based on NIR excess
\citep{sugitani02}.  
We find a spectral index of -0.3$\pm$0.1 and luminosity of
22$\pm$8\lsun.  This lower spectral index falls in the Class~II
(T-Tauri) range, and a several 10$^6$ year old object of this
luminosity falls in the 2--3\msun\ mass range of the \citet{siess} PMS
tracks.  The models consistent with the data have similar masses
(2.4--5.3\msun), and relatively low accretion rates (many disk-only
models, without an accreting envelope, fit the data well as seen in
Figure~\ref{t1}).
\citet{thompson02} derive a rather lower luminosity but had to
extrapolate from low-resolution MIR data. \citet{ma02} determine a
comparable mass to ours, 2--5M$_\sun$.  They determine an extinction
``down to the photosphere'' of $A_V$=15; 
we determine a foreground extinction of $\sim$5, and the models
that fit with $\chi_\nu^2<3.7$ have a range of circumstellar extinction from
$<$1 to $\simeq$20.  As seen in Figure~\ref{t1}, the SED would be
better defined, and the circumstellar dust distribution better
constrained, with optical measurements or a more precise 24\um flux
density (which would require higher spatial resolution to better
separate the source from nearby diffuse emission from the pillars).
If indeed this object is fairly evolved (several million years old),
then it seems unlikely that it was triggered by the actions of
NGC~6611, unless there really are 6Myr evolved stars in that open
cluster that were able to make their effects felt very early.
It is also possible that the photoablating effects 
of NGC~6611 have caused premature dispersal of the larger envelope 
around this source.

The northernmost pillar ($\Pi_1$; Fig~\ref{indiv}) is actually two head-tail
structures, or two molecular clouds \citep{pound98}.  YSOs found
half-way down the pillar obey the same paradigm of ``sources forming
at the tips of pillars'' as the sources at the three tips.  After
YSO1(P1) and YSO2(T1), two of the objects with highest estimated mass
and extinction in the pillars are located in this region: H96 EGGs E31
and E42 \citep{ma02}.  We will refer to these as P3 and T2,
respectively, following \citet{sugitani02}.  P3 (E31) is one of the
brightest objects in the pillars by 8$\;\mu$m, much brighter than its
optically-bright neighbor to the SW.  
It has a very positive spectral index 2.2$\pm$0.2 and luminosity of
15$\pm$5\lsun, which corresponds to about a 1.5\msun\ 2\e{5} year old
PMS star.  Like P1, only models with relatively massive, accreting envelopes
fit this source well; it must also be quite young. In both sources, 
we see that when an object is modeled as having a massive envelope, 
it is more difficult to constrain the mass of the accretion disk inside that 
envelope from infrared data alone.

Source T2 is similar in nature to T1:
its spectral index 0.2$\pm$0.2 is consistent with zero, and 
the majority of well-fitting models have low
accretion rate (see Fig~\ref{t2}).  Its fairly low luminosity 6$\pm$1.5\lsun\ 
corresponds to 1.5\msun\ at 10$^6$ years, and although the best-fitting models
have somewhat higher masses, the lack of optical and 24\um fluxes make it
difficult to constrain the parameters of this source beyond stating that the
probability of a very massive envelope is low.
Again, if this
object is truly so evolved, then the case for triggering by 1-2Myr old
stars in NGC~6611 is weakened.
\begin{figure}
\rplotone{f15a.eps}%P3.eps}
\rplotone{f15b.eps}%P3.mm.eps}
\caption{\label{p3} Model fits to the source P3. Plots and annotation
as Fig~\ref{p1}.}
\end{figure}

\begin{figure}
\rplotone{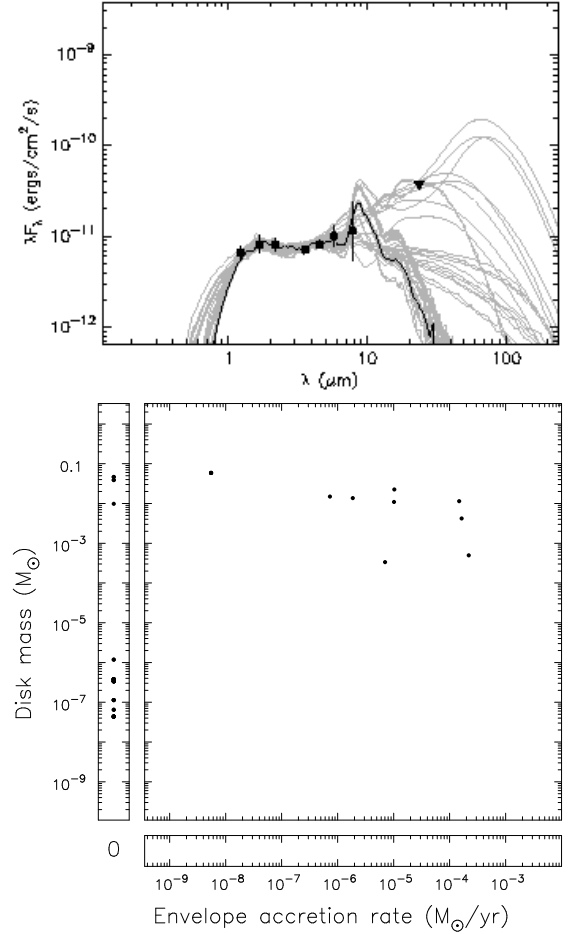}%T2.eps}
\rplotone{f16b.eps}%T2.mm.eps}
\caption{\label{t2} Model fits to the source T2. Plots and annotation
as Fig~\ref{p1}.}
\end{figure}

Two stars are located in the concave tip of $\Pi_3$, either slightly
more evolved objects, or a chance line-of-sight alignment
\citep[H96,][]{thompson02}.  We find that these objects are best-fit
with extinguished main-sequence stars, not YSOs, supporting the
theory that they are not very young objects.  
%There is a redder object
%just to the south (what we have labeled ``S3'' in Fig~\ref{indiv})
%whose SED is fit with similar parameters as P1 
%Table~\ref{ysotable2}).  
Other previously noted objects in the region, such as sources
``I'' (to the north of Pillar~1) and
``367B'' of \citet{pilbratt98} (to the east of P3) are also best fit by extinguished photospheres, and we
do not consider these strong candidates for young stellar objects.

%%%%%%%%%%%%%%%%%%%%%%%%%%%%%%%%%%%%%%%%%%%%%%%%%%%%%%%%%%%%%%%%%%%

\subsection{HH216}

HH216 is an optically visible Herbig-Haro object located just south of
the base of the three HST pillars (see Fig~\ref{indiv}).  \citet{andersen04} propose that
the driving source of HH216 is one of two very red sources located in
a structure to the south of the HST pillars, which they call pillar
\#4.  The two sources are irregularly shaped in those author's NIR
data, and are clearly detected in the IRAC data, although unresolved
with our poorer resolution.  In fact, the confused diffuse emission in
the region makes it difficult to extract a reliable flux at [8.0],
leaving the SEDs less constrained than some of the other objects which
we have discussed.  
The shape of the SED suggests quite high extinction (A$_V$ of
several tens), and the spectral indices of 1.6$\pm$0.6 and 1.1$\pm$0.5
for the northern and southern source respectively suggest youth and/or
very high extinction.  There is not clearly any associated 24\um
emission, ruling out extremely high luminosities.  The {\it Spitzer}
data are in all consistent with the solar-mass Class~I sources that
most commonly drive HH jets.
Our data do not suggest one or the other as the more likely source for
HH216.  The source apparently located more in the tip of the pillar,
at least in projection, which we call HH-N, has maser emission
detected by \citet{healy04}, a possible indicator of greater activity.

%%%%%%%%%%%%%%%%%%%%%%%%%%%%%%%%%%%%%%%%%%%%%%%%%%%%%%%%%%%%%%%%%%%

\subsection{Pillar 5}
\label{p5}

\begin{figure}[h]
\rplotone{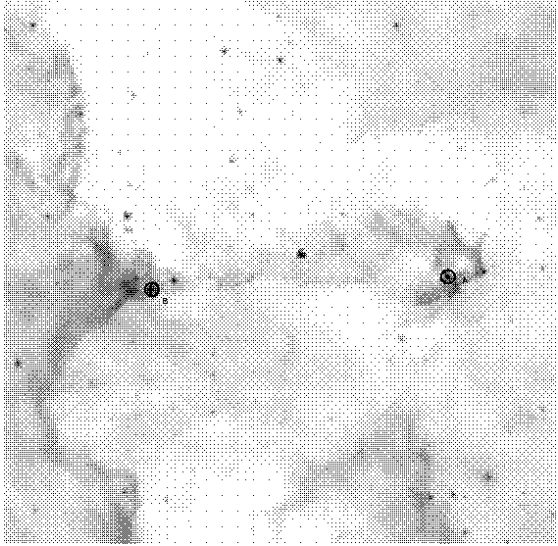}%p5_75.ps}
\caption{\label{othertrunk} IRAC [8.0] image of eastern pillar 5. Positions of 
water masers from \citet{healy04} are coincident with source ``A''.}
\end{figure}

\begin{figure}
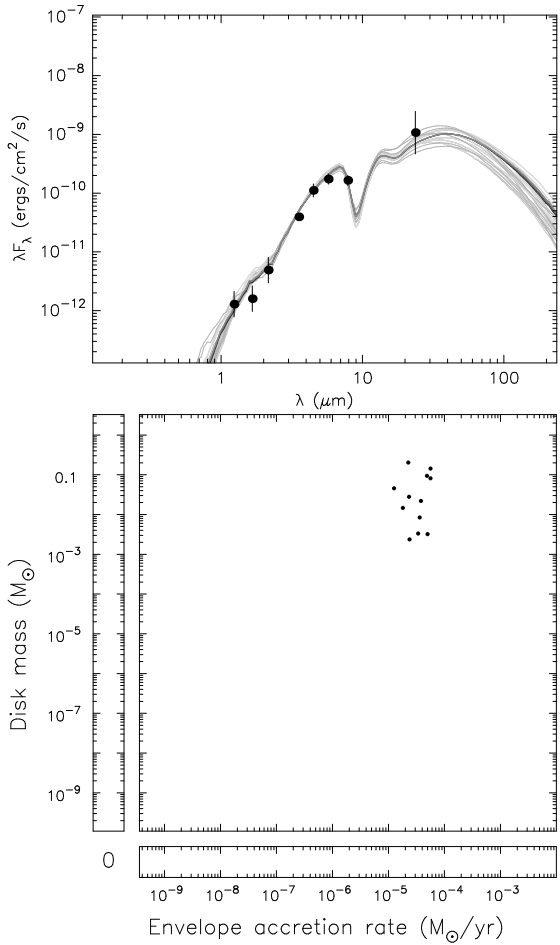

\rplotone{f18a.eps}%p5.3.eps}
\rplotone{f18b.eps}%p5.3.mm.eps}
\caption{\label{p5.3} Model fits to the pillar 5 source A. Plots and annotation
as Fig~\ref{p1}.}
\end{figure}

There are several YSO candidates in a prominent pillar to the east of
the HST pillars, near 18h19m07s -13h45m24s.  This pillar was called
pillar \#5 by \citet{healy04}, who discovered water maser emission in
several spots.  Figure~\ref{othertrunk} shows the IRAC image of pillar
5.  The masers are coincident with a bright red MIR source; in fact
the source is marginally resolved in our images, with two components
corresponding to two of the three different clusters of maser sources.
We call this source ``A'', and its SED is shown in
Figure~\ref{p5.3}.  
This source's high spectral index 2.0$\pm$0.1, high luminosity
250$\pm$100\msun, and the fact that it is a bright 24\um source all
make it a good candidate for a young intermediate mass YSO.  
At an age of 1--2\e{5}yr, that luminosity corresponds to $\sim$6\msun.  
Well-fitting models have relatively high foreground extinctions A$_V\sim$7
and quite massive envelopes of a few 10$^{-5}$\msun.
The other sources in the tips of the
``fingers'' of this pillar are also classified as YSO candidates,
mostly with modest accretion rates.
Another quite luminous young object is located 
near the base of the pillar, and is detected easily at
24\um and even at 70\um.  Source ``B'' has a high spectral index 1.7$\pm$0.1 and luminosity
200$\pm$50\lsun.

%%%%%%%%%%%%%%%%%%%%%%%%%%%%%%%%%%%%%%%%%%%%%%%%%%%%%%%%%%%%%%%%%%%

\subsection{North Bay}

Another region highlighted by \citet{healy04} for the presence of
water maser emission is to the north of the HST pillars, around
18h18m46s -13d44m30s.  There is a bright submm source here, a faint
green MIR source at the peak of the 850\um emission (not quite
coincident with a water maser), and another offset MIR source which is
exactly coincident with another water maser.  The former, faint green
source is only detected at [3.6] and [4.5].  The latter, on the ridge,
is detected only longward of [3.6], and is only well-resolved from the
diffuse emission at [4.5].  With so few extractable fluxes, it is
difficult to say anything about the nature of these sources, except
for their [3.6] and [4.5] flux densities: The faint green source has
flux densities of 1$\pm$0.5mJy and 10$\pm$3mJy, and the brighter
source on the ridge 13$\pm$5mJy and 20$\pm$3mJy in those two bands.

\begin{figure}[h]
\rplotone{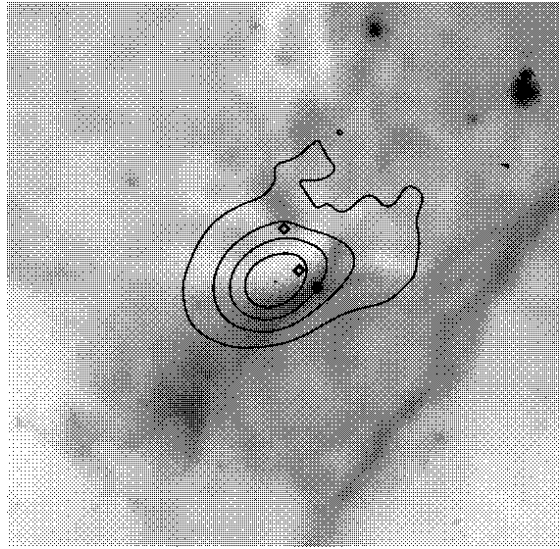}%nbay_75.ps}
\caption{\label{northbay} IRAC image of North Bay region.  SCUBA
850\um contours are superimposed (data from the SCUBA
archive, observation date Feb 18, 2002, program M01BU20), and
locations of water masers marked as diamonds \citep{healy04}.}
\end{figure}

\subsection{a new Massive Young Stellar Object}

\label{massive}
\begin{figure}[h]
\rplotone{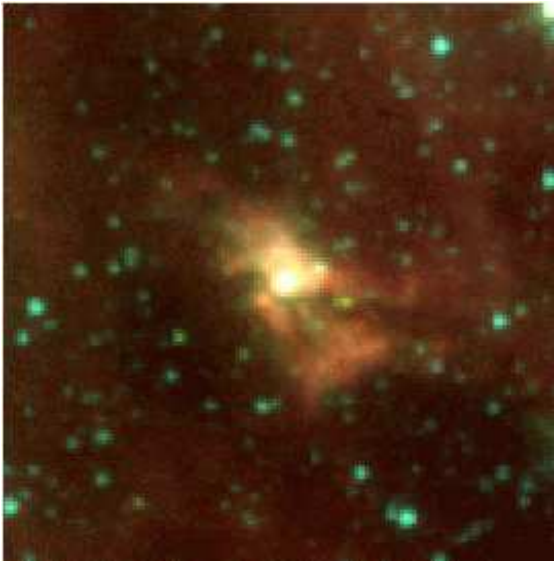}%hmpo2_75.ps}
\caption{\label{nw} IRAC image of a massive young stellar object to
the west of NGC~6611. Red, green, and blue are IRAC [3.6], [4.5], [8.0].}
\end{figure}

\begin{figure}
\rplotone{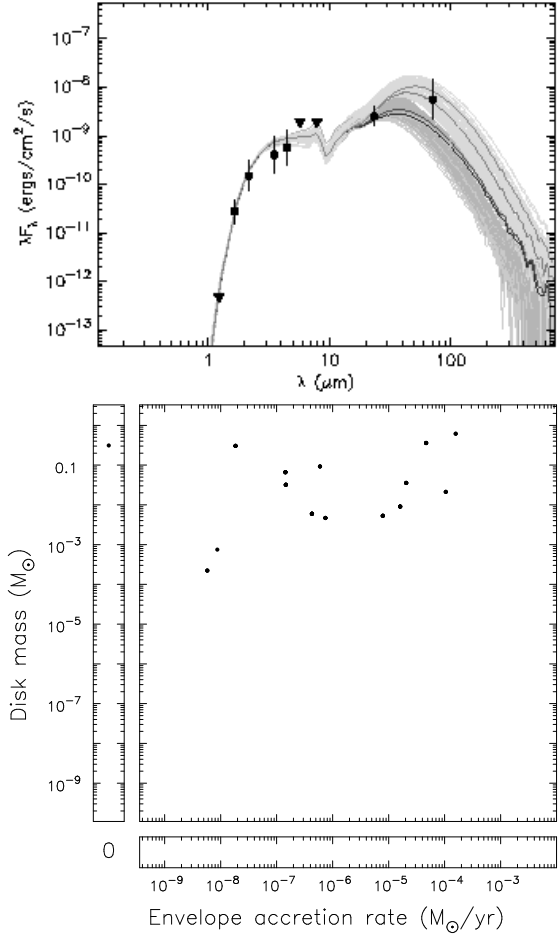}%myso.eps}
\rplotone{f21b.eps}%myso.mm.eps}
\caption{\label{mysofit} Model fits to the western massive YSO. Plots and annotation
as Fig~\ref{p1}.}
\end{figure}

The most luminous YSO which we identify in the region is labeled
``MYSO'' in Tables \ref{ysotable1} and \ref{ysotable2}, and shown in
Figure~\ref{nw}.  This YSO is an IRAS source, 18152-1346, and
it also appears to have associated water maser emission \citep{braz},
another sign of massive star formation. 
The shorter wavelengths (2MASS, IRAC [3.6], [4.5]) show a bright
central point source surrounded by either a small cluster, not
resolved into individual stars, or extended diffuse emission as might
arise in an outflow.  The IRAC colors do not show a particular excess
at [4.5], as sometimes occurs in sources believed to have outflows
(and emission in that band from CO and H$_2$ lines).  Precise
photometry of this source is made difficult by the extended nature at
short wavelengths, and that it is saturated at [8.0] and [24], and
marginally saturated at [5.8].  We used a combination of PSF wing
fitting, aperture photometry, and comparison with the MSX [8] and [21]
fluxes to derive the source's SED shown in Figure~\ref{mysofit}.  The
source's spectral index 1$\pm$0.3 and luminosity 1000$\pm$300\lsun\ suggest
that it is young, and may have a mass (from PMS tracks) of about
8\msun.  Examination and fitting of the SED shows likely PAH emission
from the source, increasing the fluxes at [5.8] and [8.0].  PAH
emission is expected from the surface of the accretion disk, or
interior of an outflow cavity in a flattened envelope, for such more
massive sources with stronger radiation fields.  We modeled the source
with the measured fluxes in those bands, and also treating them as
upper limits (since our model grid does not include PAH emission), and
this difference did not greatly change the results. The source is not
detected at J, so it must still be very embedded in its molecular
cloud -- indeed, the best-fitting models have A$_V\simeq$30.  Those
models also have quite massive disks and envelopes, with accretion
rates of a few times 10$^{-5}$\msun/yr.  However, considering that this
is an intermediate to massive star, that accretion rate does not
necessarily imply extreme youth.
Clearly, followup with spectroscopy and higher resolution photometry
at longer wavelengths would be interesting to clarify the nature of
this source, which may be responsible for some of the extra heating in
that part of the molecular cloud seen in Figure~\ref{dusttemp}.

% iras 14.2 64.9 549.4 1169

%%%%%%%%%%%%%%%%%%%%%%%%%%%%%%%%%%%%%%%%%%%%%%%%%%%%%%%%%%%%%%%%%%%
%%%%%%%%%%%%%%%%%%%%%%%%%%%%%%%%%%%%%%%%%%%%%%%%%%%%%%%%%%%%%%%%%%%

%\clearpage
\section{Conclusions}
\label{conclusions}

We analyze {\it Spitzer} IRAC/GLIMPSE images of the M16 region.  The
diffuse emission in the long-wavelength IRAC bands (most likely PAH
emission) traces the bubble blown by the young open cluster NGC~6611,
and highlights the famous HST star-forming pillars.  We use a large
grid of model spectral energy distributions of young stellar objects
to classify all the point sources in the area ($\leq$0.45$^\circ$ of $l$=17.0,
$b$=0.8), and select candidate YSOs.  The source classification
determined using our model grid is comparable to classification based
on IRAC colors \citep[e.g][]{allen,megeath1}.  Simultaneous use of all
photometric information between 1 and 25\um provides additional
insight into physical parameters like envelope accretion rate and disk
mass, and how tightly the data can constrain the circumstellar dust
distribution.  We
particularly highlight sources that can only be fit well by models with massive envelopes, and are thus likely very young.
The evolutionary classification possible using MIR data is not always
the same as results based on NIR color excess alone.

The distribution of young objects in the region is consistent with a
picture of moderate distributed star formation, but no particularly
vigorous activity triggered by the optical cluster NGC~6611.  We
confirm the youthful nature of that cluster, previously quantified as
a large fraction of L-band (3.5$\;\mu$m) excess sources \citep[85\%][]{oliviera},
and a large number of emission-line stars \citep[52, ][]{dewinter97}.
Nearly all ($>$95\%) of the sources in the NGC~6611 cluster that are
detected in at least 4 of the 7 bands (J,H,K,[3.6],[4.5],[5.8],[8.0])
are YSO candidates.
We identify a second cluster of YSOs to the northwest, apparently in a
concentration of dense gas, and a cluster of YSOs at the tip of pillar
5 to the east.
Examination of the two-point correlation function reveals that the 
YSO candidates are highly clustered.  The nearest neighbor distribution 
suggests a characteristic size scale $\sim$0.1pc, which corresponds
to the Jeans mass in the clumps at the heads of the pillars.
Mid-IR imaging allows us to relate the YSO population to the physical
conditions in their natal cloud and to paint a consistent picture of
the star formation in M16.  A newly discovered intermediate to massive YSO may be
responsible for additional heating of the molecular cloud west of
NGC~6611.

Finally, we present a detailed examination of selected sources (for
example in the HST pillars).  Analysis of 
the 2--24\um spectral index, bolometric luminosity, and 
all of the well-fitting
model SEDs allows us to constrain the evolutionary states, stellar and disk masses,
and accretion rates, in some cases much more tightly than was possible
from NIR data alone.  Several sources in the trunks are very young/unevolved
%($\lesssim$10$^5$yr), 
and it is possible that these sources were
triggered by the compressive action of NGC~6611.  The small ages of
those sources could even be consistent with the very short destruction
timescale for the trunks by velocity gradients
\citep[10$^5$yr][]{pound98}.  On the other hand, some of the other
sources in the trunks do not have massive envelopes;  
these may be more evolved, consistent with Class~II or T-Tauri sources, 
or perhaps photoablation from NGC~6611 has accelerated the 
dispersal of their envelopes (both are located at pillar tips).
If these sources are more than a few $\sim$10$^6$yr old,
it is difficult to understand how they could have been formed by
triggering action of 1-2Myr old stars in NGC~6611.  
We also do not see any particular pattern in evolutionary state, with younger sources 
farther from NGC6611, as might occur in a situation dominated by triggering 
\citep[although we note that a pattern might be hidden by the 3D structure of the pillars region as revealed by][]{pound98}.
We favor the interpretation that YSOs formed spontaneously in dense parts of
the cloud, and then these densest cores have persisted because they
are more robust to photo-destruction than surrounding material; 
YSOs are found in exposed pillars, but were not necessarily
formed as a result of that exposure.

\section{Acknowledgements}

This project makes use of {\it Spitzer}'s GLIMPSE and MIPSGAL survey
data.  Support for the work was provided by NASA through Spitzer
Legacy Science Program contracts 1224988 (BAW) and 1224653 (EC, BB, MM,
CW, RI) and the Spitzer Fellowship Program (RI).  BAE also acknowledges 
NASA's Astrophysics Theory Program (NNG05GH35G).
This publication makes use of data products from the Two Micron All
Sky Survey, which is a joint project of the University of
Massachusetts and the Infrared Processing and Analysis
Center/California Institute of Technology, funded by the National
Aeronautics and Space Administration and the National Science
Foundation
This research has made use of NASA's Astrophysics Data System Bibliographic Services.

\clearpage
\begin{landscape}
\begin{deluxetable}{lcc|cccccccc|cc}
%\rotate
\setlength{\tabcolsep}{1mm}
\tablewidth{0pt}
\tabletypesize{\scriptsize}
\tablecaption{Specific YSO candidates: observed parameters}
\tablehead{
\colhead{name} & \colhead{R.A.} & \colhead{Dec.} & 
\colhead{J} & \colhead{H} & \colhead{K$_s$} &
\colhead{[3.6]} & \colhead{[4.5]} & \colhead{[5.8]} & \colhead{[8.0]} & \colhead{[24]} &
\colhead{spectral index} & \colhead{log(L$_\star$)} \\
& \colhead{(J2000)} & \colhead{(J2000)} & 
\colhead{[mJy]} & \colhead{[mJy]} & \colhead{[mJy]} & \colhead{[mJy]} &
\colhead{[mJy]} & \colhead{[mJy]} & \colhead{[mJy]} & \colhead{[mJy]} & 
2--24\um & \colhead{log(L$_\sun$)}
}
\startdata
P1 & 18h16m00.1s & -13d50m11s & 2.5$\pm$0.5 & 6.6$\pm$1.0 & 12.$\pm$2.0 & 
26.$\pm$4.0 & 50.$\pm$10. & 45.$\pm$30. & 40.$\pm$30. & 2000$\pm$1000 &
1.0$\pm$0.2 & 44$\pm$16 
\\
P2 & 18h18m51.5s & -13d49m13s & 0.4$\pm$0.25 & 1.0$\pm$0.3 & 2.7$\pm$0.5 &
4.3$\pm$0.5 & 5.9$\pm$0.5 & 10.6$\pm$2.0 & 70$\pm$60 & $<$420\tablenotemark{a} &
0.3$\pm$0.2 & $\gtrsim$5.3\tablenotemark{b}
\\
T1 & 18h15m58.5s & -13d51m06s & 6.4$\pm$0.7 & 12$\pm$1.2 & 19$\pm$2.0 & 
28$\pm$8.3 & 29$\pm$4.4 & 43$\pm$8.6 & 40$\pm$10 & 560$\pm$400 &
-0.3$\pm$0.1 & 22$\pm$8 
\\
P3 & 18h18m52.3s & -13d49m38s & 1.0$\pm$0.7 & 1.5$\pm$0.5 & 2.2$\pm$0.5 & 
4.8$\pm$0.5 & 17$\pm$2.0 & 46$\pm$5.0 & 70$\pm$40 & 400$\pm$200 &
2.3$\pm$0.2 & 15$\pm$5 
\\
T2 & 18h18m53.2s & -13d49m57s & 2.8$\pm$0.5 & 4.6$\pm$1.0 & 5.9$\pm$1.0 &
8.6$\pm$0.9 & 12$\pm$1.2 & 20$\pm$5.0 & 40$\pm$30 & $<$300\tablenotemark{a} &
0.2$\pm$0.2 & 6.0$\pm$1.5 
\\
HH-N & 18h18m58.9s & -13d52m48s & 0.1$\pm$0.09 & 0.3$\pm$0.2 & 1.0$\pm$0.5 &
3.0$\pm$0.5 & 6.3$\pm$1.0 & 6.0$\pm$4.0 & $<$41\tablenotemark{a} & $<$310\tablenotemark{a}&
1.6$\pm$0.6 & $\gtrsim$1.4\tablenotemark{b}
\\
HH-S & 18h18m59.4s & -13d52m57s & 0.05$\pm$0.04 & 0.2$\pm$0.1 & 1.3$\pm$0.5 &
1.7$\pm$0.5 & 5.7$\pm$6.0 & 6.0$\pm$2.0 & 40$\pm$30 & $<$270\tablenotemark{a} &
1.1$\pm$0.5 & $\gtrsim$3.0\tablenotemark{b} 
\\\hline
%P5 B & 18h19m07.2s & -13d44m35s &  0.5$\pm$0.3 & 2.3$\pm$0.4 & 7.0$\pm$1.0 &
%12.$\pm$2. & 7.$\pm$2. & 15.$\pm$4. & 40.$\pm$20. & &
%25$\pm$4 & 1.6$\pm$0.5 
% possibly PAH contaminated!
%\\
% p5.3:
P5 A & 18h19m07.2s & -13h45m24s & 0.6$\pm$0.3 & 1.0$\pm$0.5 & 4.0$\pm$2.0 &
47$\pm$4.7 & 170$\pm$44 & 340$\pm$60 & 440$\pm$80 & 12000$\pm$10000 &
2.0$\pm$0.1 & 250$\pm$100 
\\
%P5 C & 18h19m16.23s & -13d45m04.1s & $<$1        & 8.5$\pm$1.0 & 80$\pm$10  & 400$\pm$50   & 570$\pm$50   & 750$\pm$50  & 800$\pm$100  & &
%&
%\\
% p5base:
P5 B & 18h19m25.33s & -13d45s35.5s & 0.5$\pm$0.3 & 3.0$\pm$0.4 & 24$\pm$2.4 & 200$\pm$21   & 320$\pm$150  &  900$\pm$170  & 1000$\pm$800 & 2700$\pm$1000 &
1.7$\pm$0.1 & 200$\pm$50 
\\
MYSO & 18h18m08.6s  & -13d45m06s   & $<$0.2      & 18$\pm$10   & 140$\pm$100& 7000$\pm$500 & 1200$\pm$800 & 3700$\pm$1500 & 5000$\pm$3000 & 22000$\pm$10000 & 
1.0$\pm$0.3 & 1000$\pm$300 
\\
\enddata
\label{ysotable1}
\tablenotetext{a}{3$\sigma$ upper limit}
\tablenotetext{b}{Without a measured flux density at 24$\;\mu$m, 
$\Sigma\nu F_\nu$ 
is certainly lower than the true bolometric luminosity.  Since
we do not include flux densities longward of MIPS-24$\;\mu$m, all of the
luminosities in this table may be low - this is reflected in the
luminosities and masses of well-fitting models being somewhat
(10-100\%) higher.}
\end{deluxetable}

%%%% the lower version had dchi=3, and figures with d(cpd)=1.5
% we change to E1 (cpd<1) for both?

\begin{deluxetable}{c|ccc|cccc|ccc|ccc}
\setlength{\tabcolsep}{1mm}
\tablewidth{0pt}
\tabletypesize{\footnotesize}
\tablecaption{Specific YSO candidates: derived parameters}
\tablehead{
\colhead{name} &
\multicolumn{3}{c}{A$_V$} &
\multicolumn{4}{|c}{M$_\star$ [M$_\sun$]} & 
\multicolumn{3}{|c}{log(\.{M} [M$_\sun$/yr])} & 
\multicolumn{3}{|c}{log(M$_{disk}$ [M$_\sun$])} \\
& min & ave\tablenotemark{a} & max & estimate\tablenotemark{b} & min & ave & max & min & ave & max & min & ave & max
}
\startdata
%     AV             mass                   mdot              mdisk
P1   & 2.0\tablenotemark{c}
          & 4.4& 7.0& 3 & 2.9 & 4.6 & 6.5 & -4.8& -4.5& -4.2& -3.1& -1.7& -0.5\\
P2   & 2.0& 14 & 19 & 1 & 0.4 & 1.5 & 3.7 & --\tablenotemark{d}  
                                                & -5.2& -3.5& -4.0& -2.5& -1.3\\
% P3 and T1 we have to do the plot and fit with D2.7 to reflect that the end data points
% are pretty ratty.
T1   & 2.0& 5.0& 5.7&2.5& 2.6 & 3.2 & 5.3 & --  & -6.1& -4.8& -3.6& -2.3& -0.8\\
P3   & 2.0& 5.9& 9.1&1.5& 0.9 & 1.8 & 5.7 & -5.3& -4.3& -3.2& -4.0& -2.9& -1.2\\
T2   & 2.0& 8.4& 11 &1.5& 0.2 & 3.2 & 5.5 & --  & -6.3& -3.7& -7.4& -5.4& -1.2\\
%\hline
%HH-N& 2.0& 12 & 26 &   & 0.1 & 1.3 & 8.0 & --  & -4.9& -2.5& -8.1& -2.9& -0.2\\
%HH-S& 2.0& 21 & 31 &   & 0.1 & 1.7 & 6.0 & --  & -5.5& -3.5& -7.4& -3.6& -1.0\\
%p5.3:
P5A  & 2.3& 6.8& 8.8 & 6 & 0.5 & 3.1 & 6.5 & -4.9& -4.5& -4.3& -2.6& -1.6& -0.7\\
% ??:
%P5C  &&&&&&&&&&&&&\\
% p5base not very well fit
%P5D & 3.4& 7.0& 9.4&   & 6.2 & 6.9 & 10  & --  & -8.5& -3.4& -3.2& -1.8& -1.0\\
MYSO & 27 & 31 & 33 & 8 & 7.7 & 8.8 & 9.8 & --  & -6.2& -3.8& -3.6& -1.6& -0.2\\
\enddata
\label{ysotable2}
\tablenotetext{a}{For each parameter, the minimum and maximum value
for any model which fits the data better than $\chi_\nu^2<3.7$
(probability of 0.001) is given, along with the average weighted by
the probability exp($-\chi^2$/2).}
\tablenotetext{b}{Mass estimated from an age consistent with the
2--24\um spectral index, and the measured bolometric luminosity, using
PMS tracks from \citet{siess}. This method is not expected to be more
accurate than a factor of 50\% in mass.}
\tablenotetext{c}{\ Line-of-sight extinction was restricted to be greater than 
$A_V$=2, since \citet{hillenbrand93} determined that the extinction to NGC6611
is $A_V$=3.1, and that cluster is in one of the less embedded parts of M16.}
\tablenotetext{d}{Some disk-only models, with no accreting envelope,
fit the data.}
\end{deluxetable}
\clearpage
\end{landscape}

\end{document}